%% file: main.tex
\documentclass[twocolumn, twocolappendix]{aastex631}

\usepackage{float}
\begin{document}

\title{Spatially Resolving the Star Formation Histories of Three Nearby Nuclear Star Clusters}

\author[0000-0002-7064-3867]{Christian H. Hannah}
\affiliation{Department of Physics and Astronomy, University of Utah\\ 115 South 1400 East, Salt Lake City, Utah 84112, USA}

\author[0000-0003-0248-5470]{Anil C. Seth}
\affiliation{Department of Physics and Astronomy, University of Utah\\ 115 South 1400 East, Salt Lake City, Utah 84112, USA}

\author[0000-0002-5678-1008]{Dieu D. Nguyen}
\affiliation{National Astronomical Observatory of Japan (NAOJ), National Institute of Natural Sciences (NINS)\\ 2-21-1 Osawa, Mitaka, Tokyo 181-8588, Japan}
\affiliation{Department of Physics and Astronomy, University of Utah\\ 115 South 1400 East, Salt Lake City, Utah 84112, USA}

\author[0000-0003-0234-3376]{Antoine Dumont}
\affiliation{Department of Physics and Astronomy, University of Utah\\ 115 South 1400 East, Salt Lake City, Utah 84112, USA}

\author[0000-0002-6072-6669]{Nikolay Kacharov}
\affiliation{Leibniz-Institut f\"{u}r Astrophysik Potsdam (AIP), An der Sternwarte 16, D-14482 Potsdam, Germany}

\author[0000-0002-6922-2598]{Nadine Neumayer}
\affiliation{Max Planck Institut f\"{u}r Astronomie, K\"{o}nigstuhl 17, D-69117 Heidelberg, Germany}

\author{Mark den Brok}
\affiliation{Department of Physics, ETH Zürich, Wolfgang-Pauli-Str 27, CH-8093 Zürich, Switzerland}
\affiliation{Leibniz-Institut f\"{u}r Astrophysik Potsdam (AIP), An der Sternwarte 16, D-14482 Potsdam, Germany}


\begin{abstract}

The formation of nuclear star clusters (NSCs) remains an open question. In this work, we use spatially resolved \textit{HST}/STIS spectroscopic observations of three nearby NSCs (hosted by NGC~5102, NGC~5206, and NGC~205) to constrain their formation histories by exploring radial variations of the stellar populations within each cluster. Utilizing full-spectrum fitting, we find substantial age and metallicity gradients within the central 0$\farcs$9 (16~pc) of the NSC in NGC~5102 where populations near the center are young/metal-rich (age $\sim$400~Myr and [M/H] $\sim$-0.4) and become older/metal-poor at larger radii (mean age $\sim$1~Gyr and mean [M/H] $\sim$-1.6 in the radial range [0$\farcs$3, 0$\farcs$9]). This behavior suggests that the young/metal-rich population at the center was formed from a period of \textit{in situ} formation, while the older/metal-poor populations were likely formed by inspiraled globular clusters. The two broad populations observed in the NGC~5102 NSC (young/metal-rich and old/metal-poor) appear to be linked to the transition between the two morphological components of the NSC derived from the surface-brightness profile in \citet{Nguyen2018}. The radial ranges explored in NGC~5206 and NGC~205 were much smaller due to poor data quality; in NGC~5206 we find a similar metallicity gradient to NGC~5102 (but with much lower significance), while the data for NGC~205 is too poor to reach any conclusions. Overall, this data highlights the links between the morphological and stellar population complexity of NSCs and their formation mechanisms.

\end{abstract}



\section{Introduction} \label{sec:introduction}
\par Nuclear star clusters (NSCs) are the densest stellar systems in the Universe and are present in nearly all galactic centers regardless of mass and morphology \citep[see review by][]{Neumayer2020}. These clusters are usually more massive than globular clusters (GCs) with masses ranging from $\sim$10$^5$~M$_\odot$ to $\sim$10$^8$~M$_\odot$ and effective radii in the range 1-40~pc with most below 10~pc. In addition to living in galactic centers, tidal stripping of galaxies can leave behind NSCs that may hide amongst the most massive globular clusters \citep[e.g.][]{Pfeffer2013,Pfeffer2021}.  Their extreme densities lead to unique dynamics such as the formation of compact binaries that lead to LIGO sources \citep[e.g.][]{Antonini2016}.  There is also evidence that stars within NSCs are the primary contributors to tidal disruption events \citep{Pfister2020}, which may be used in the future to confirm the existence of elusive intermediate-mass black holes.  Despite the clear significance of these stellar clusters, their formation is not yet fully understood. 

\input{GAL_PROPERTY_TABLE}
\vspace{-24pt}

\par Currently, two dominant theories exist for the formation of NSCs. The cluster inspiral formation mechanism involves building the NSC from massive star clusters (i.e.~GCs) that settle to the center of the galaxy due to dynamical friction \citep[e.g.][]{Tremaine1975,Capuzzo-Dolcetta1993, Miocchi2006,Antonini2013,Gnedin2014}.  Recent observational and theoretical results provide strong evidence for the importance of dynamical friction inspiral in forming the NSCs of dwarf galaxies ($M_\star < 10^9 \rm{M}_\odot$).  First, the fraction of dwarf galaxies with NSCs tracks the fraction with globular clusters very closely \citep{Sanchez-Janssen2019, Carlsten2021}.  
Second, in dwarf galaxies the NSC mass grows with the galaxy stellar mass roughly as $\sqrt{M_\star}$ \citep{denBrok2014,Sanchez-Janssen2019}, 
matching expectations from theory \citep{Gnedin2014}. Third, spectroscopic observations of NSCs in dwarf galaxies suggest they are typically more metal-poor than their surrounding host galaxies \citep{Koleva2009,Paudel2011,Spengler2017,Neumayer2020,Johnston2020,Fahrion2021}.  Metal-poor NSCs are expected in the case of cluster inspiral because the majority of Milky Way GCs \citep{Harris1996} and almost all of those in dwarf galaxies \citep{Peng2006} are metal-poor.   Using simulations, \citet{Pfeffer2018} find that most massive clusters formed in early cosmic epochs (redshifts z$\gtrsim$1), resulting in a high fraction of these clusters observed today being old and metal-poor \citep{Lamers2017}. Given this background, the presence of old and metal-poor stars in NSCs provides evidence for cluster inspiral.

\par Another possible formation mechanism for NSCs, known as \textit{in situ} formation, involves the stars forming from the accretion of gas into the galaxy's nucleus \citep[e.g.][]{Milosavljevic2004}. In this case, the resulting stellar population is expected to be rather young and metal-rich when compared to the surrounding population. \textit{In situ} formation could also be indicated by the presence of strong rotation and flattening (i.e. low $b/a$ ratios) in a younger component of an NSC's stellar population \citep[e.g.][]{Seth2006, Hartmann2011}.
Similar populations could theoretically result from the inspiral of young stars/clusters that formed just outside the nucleus \citep[e.g.][]{Gerhard2001, Agarwal2011, Neumayer2011, Hartmann2011, Antonini2014}.  However, these clusters will typically disrupt at larger radii \citep{Antonini2013}, while a more compact population of young stars, like those found in the Milky Way and other nearby NSCs in late-type galaxies \citep{Feldmeier-Krause2015, Carson2015}, argues in favor of \textit{in situ} formation.

\par Many attempts have been made to constrain the formation histories of NSCs in a variety of host galaxies and environments. Integrated spectroscopy of NSCs has been used to study the stellar populations of the clusters, providing important hints to their formation mechanisms.  This includes work on nearby bulgeless spirals by \citet{Walcher2006} and \citet{Seth2006}, which show that all the NSCs studied experienced some \textit{in situ} star formation within the past 100~Myr. Similarly, the results of \citet{Rossa2006} show recent \textit{in situ} formation in the NSCs of 40 early- and late-type spiral galaxies. More recent work by \citet{Kacharov2018} found evidence for both old and young components in late-type NSCs, suggesting that stars resulting from globular cluster inspiral may also be present in these clusters. Similar results were found for 12 early- and late-type nucleated dwarf galaxies ($<$10$^9$~M$_\odot$) by \citet{Johnston2020} providing further evidence for both \textit{in situ} and cluster inspiral formation in NSCs.

\par In addition to stellar ages, metallicity measurements of NSCs have also been used to deduce possible formation scenarios. Most of these studies have focused on early-type galaxies, where the older stellar populations make it possible to use line index measurements to measure metallicities \citep{Koleva2009,Paudel2011}. Overall, these results show that in higher mass galaxies ($\gtrsim$10$^9$~M$_\odot$), NSCs are typically more metal-rich than their host galaxies \citep{Neumayer2020,Fahrion2021}.  This, combined with the flattening observed in some of these same clusters argues for \textit{in situ} star formation \citep{Spengler2017}.  However, at lower masses, the clusters are often more metal-poor when compared to their host galaxies, suggesting cluster inspiral is increasingly important at lower galaxy masses ($\lesssim$10$^9$~M$_\odot$) \citep{Paudel2011,Neumayer2020,Fahrion2021}. \citet{Fahrion2021} find stellar populations consistent with both formation mechanisms occurring simultaneously in early-type galaxies and that the dominant formation channel for NSCs is related to the host galaxy mass by investigating the NSC formation histories of 25 nucleated galaxies with masses between $\sim$10$^8$ and 10$^{10.5}$~M$_\odot$. These results provide additional evidence for the suggestion that there is a transition at a galaxy mass of $\sim$10$^9$~M$_\odot$  between cluster inspiral formation at lower galaxy masses and \textit{in situ} formation at higher galaxy masses \citep{Turner2012,Sanchez-Janssen2019,Neumayer2020}.  The age and metallicity measurements discussed here are all integrated quantities for their NSCs. However, it is clear that NSCs have a complex internal structure. For instance in NGC~404, \citet{Nguyen2017} found that the counter-rotating core of the NSC is dominated by $\sim$1~Gyr old stars, while stars at larger radii within the NSC are older.  

\par The most detailed view we have of an NSC is in the Sagittarius dwarf galaxy, which hosts the NSC known as M54.  Resolved stars within the cluster have been used to construct detailed color-magnitude diagrams (CMDs) \citep[e.g.][]{Siegel2007,Alfaro-Cuello2019}. Most recently, \cite{Alfaro-Cuello2019} and \citet{Alfaro-Cuello2020} have used metallicitiy measurements for individual stars combined with isochrone fits to the CMD of the cluster to distinguish three populations: young/metal-rich (YMR), intermediate-age/metal-rich (IMR), and old/metal-poor (OMP). The OMP population was found to have a significant metallicity spread and almost no rotation suggesting the inspiral of multiple GCs. The YMR population on the other hand shows strong rotation and appears to result from \textit{in situ} formation.  

\par Clearly, the ability to resolve the stellar populations within an NSC allows for a more detailed investigation of its formation history. For this reason, we use spatially resolved spectroscopic observations (0$\farcs$05 pix$^{-1}$) of our sample galaxies to probe the formation histories of their NSCs. Our investigation focuses on three nearby, early-type dwarf galaxies: NGC 5102, NGC 5206, and NGC 205. NGC 5102 and NGC 5206 are both members of the Cen A galaxy group, while NGC 205 is a satellite galaxy of Andromeda. General properties of these galaxies and their NSCs are outlined in Table~\ref{tab1}. 
For a more detailed description of these galaxies and previous measurements, see \citet{Nguyen2018}. All three galaxies have masses between 10$^9$ and 10$^{10}$~M$_\odot$, a mass range where \citet{Fahrion2021} find stellar population evidence for a mix in star formation mechanisms. Although the stellar populations of our target galaxies and their NSCs have been previously studied \citep[e.g.][]{Monaco2009,Mitzkus2017,Kacharov2018}, these studies lack the spatial resolution of the spectroscopic data utilized in this work. To further verify our analysis, direct comparisons between our findings 
and those of \citet{Kacharov2018} and \citet{Mitzkus2017} are made in Section~\ref{sec:comparisons}. We primarily focus on results for NGC~5102 in this paper due to poor data quality of the spectra for NGC~5206 and NGC~205 (see Section~\ref{sec:otherresults} for more details).

\section{Data \& Reduction} \label{sec:data}
\par Each galaxy in this study has been observed with the \textit{Hubble Space Telescope's} (\textit{HST's}) Space Telescope Imaging Spectrograph (STIS) resulting in spatially resolved, long-slit spectroscopic data \citep{Nguyen2019}. During each observation, the G430L grating was used with the 52$''$ x 0$\farcs$1 slit. The slit was oriented with the following angular offsets from the NSC major axes: $\sim$11$^{\circ}$ for NGC~5102, $\sim$21$^{\circ}$ for NGC~5206, and $\sim$10$^{\circ}$ for NGC~205. The resulting 2-dimensional data set provides the flux as a function of both wavelength and spatial location. The G430L grating provides a spectrum over the wavelength range [2900~\AA - 5700~\AA] with an average dispersion of 2.73 $\textrm{\AA/pix}$ and a spatial sampling of 0.05078~$''$/pix.

\par We downloaded multiple reduced and rectified observations of each galaxy from the \textit{HST}/Hubble Legacy Archive as follows: 9 exposures (933 seconds each) for NGC 5102, 5 exposures (946 seconds each) for NGC 205, and 7 exposures (1924 seconds each) for NGC 5206. For each exposure, the sources were placed in the E1 aperture position to reduce charge transfer efficiency effects. Furthermore, the telescope was dithered between exposures for a single galaxy to reduce systematic errors such as spatially varying chip sensitivity, small-scale detector defects, and cosmic rays. As a result, the images were shifted to align the center spectra during combination. 
\par The spectrum corresponding to the center of the galaxy in each image was determined by taking the median flux across all wavelengths and finding the maximum along the spatial axis. The central spectra of each exposure were then visually inspected. In some images, the spatial index corresponding to the NSC center was shifted by a single pixel to maximize consistency between the between the spectral shape and features in the individual exposures.

\begin{figure}[htp]
    \begin{flushleft}
        \begin{minipage}{\linewidth}
            \includegraphics[width=\linewidth]{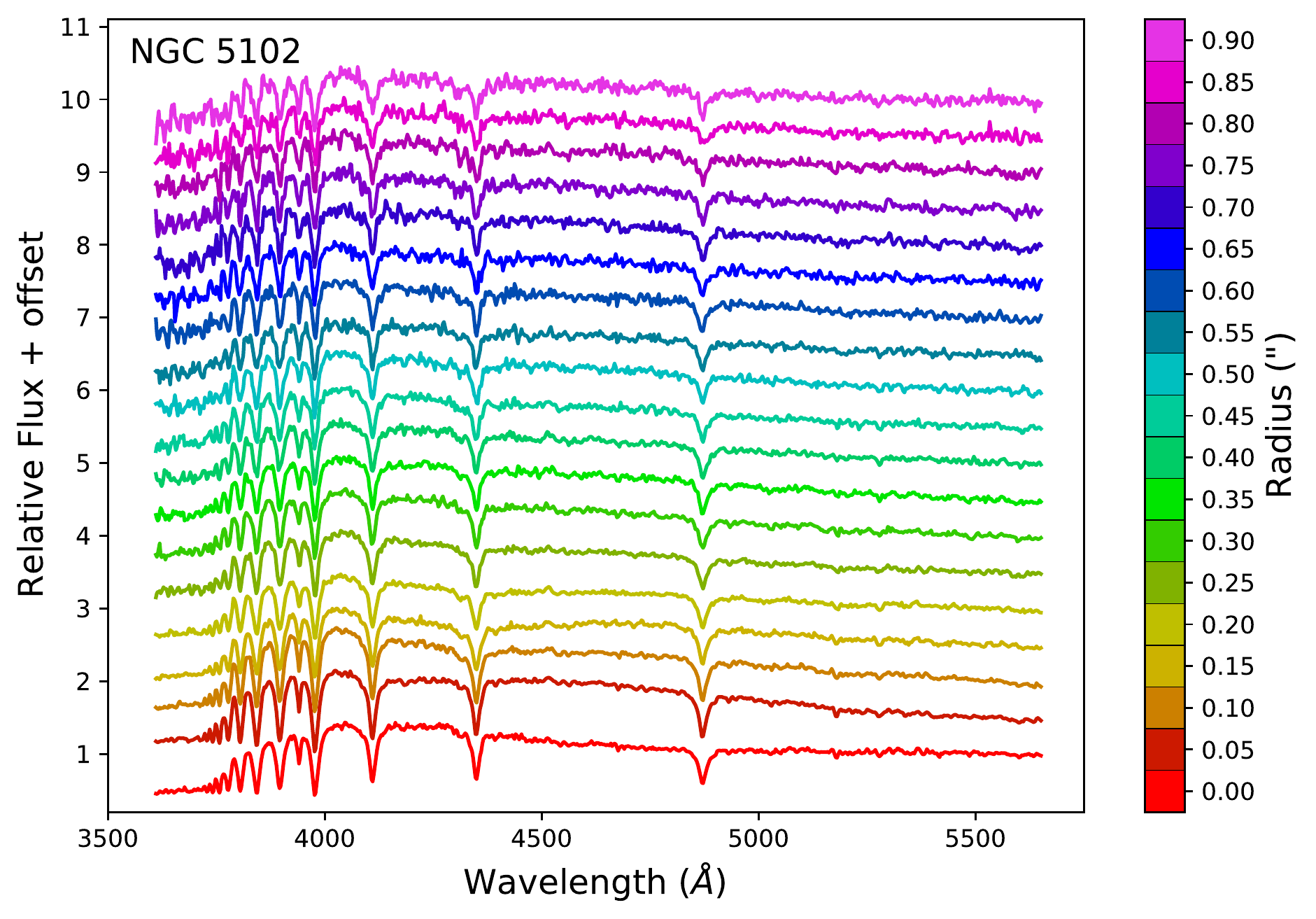}
        \end{minipage}
        \hfill
        \begin{minipage}{\linewidth}
            \includegraphics[width=\linewidth]{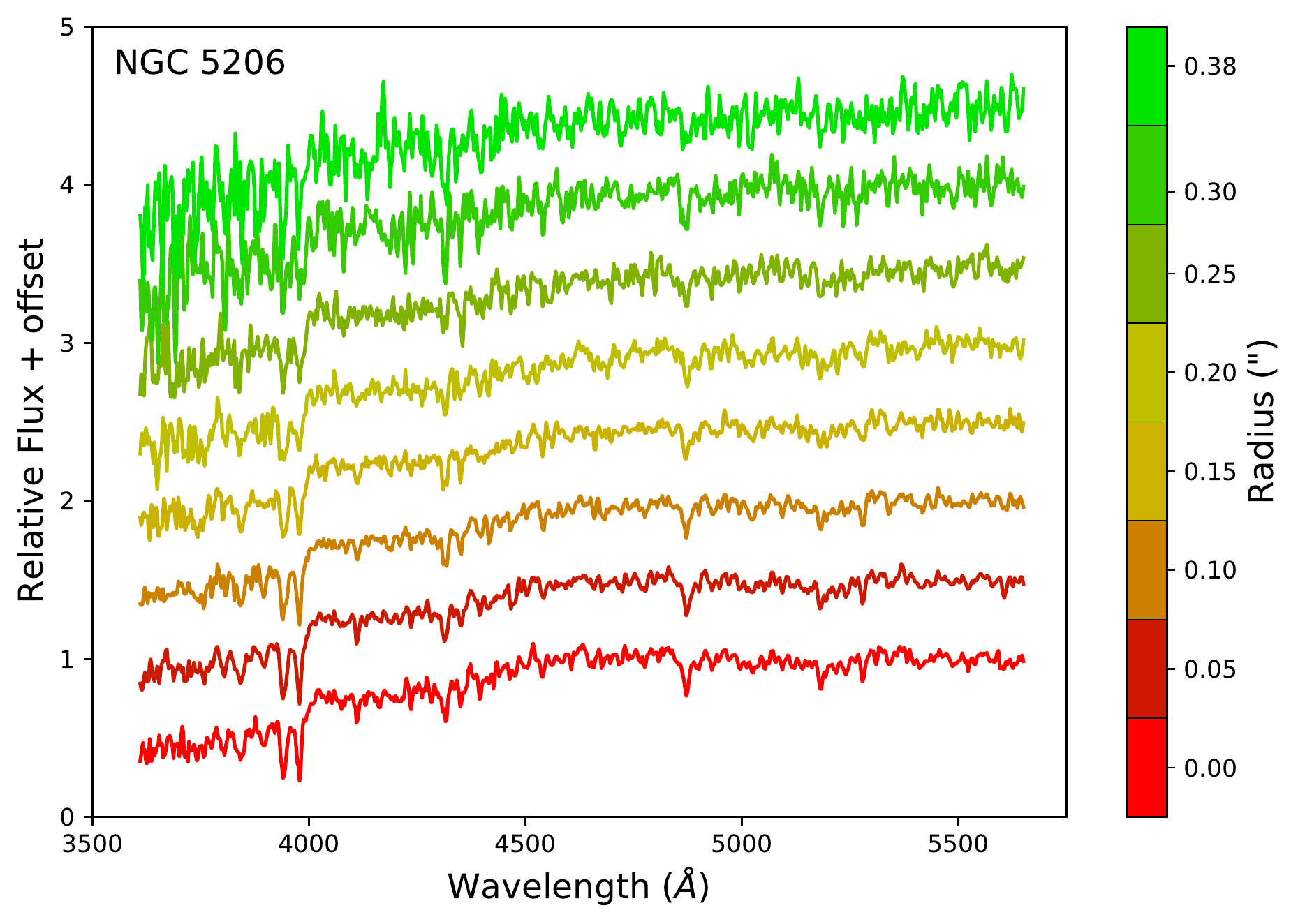}
        \end{minipage}
        \hfill
        \begin{minipage}{\linewidth}
            \includegraphics[width=\linewidth]{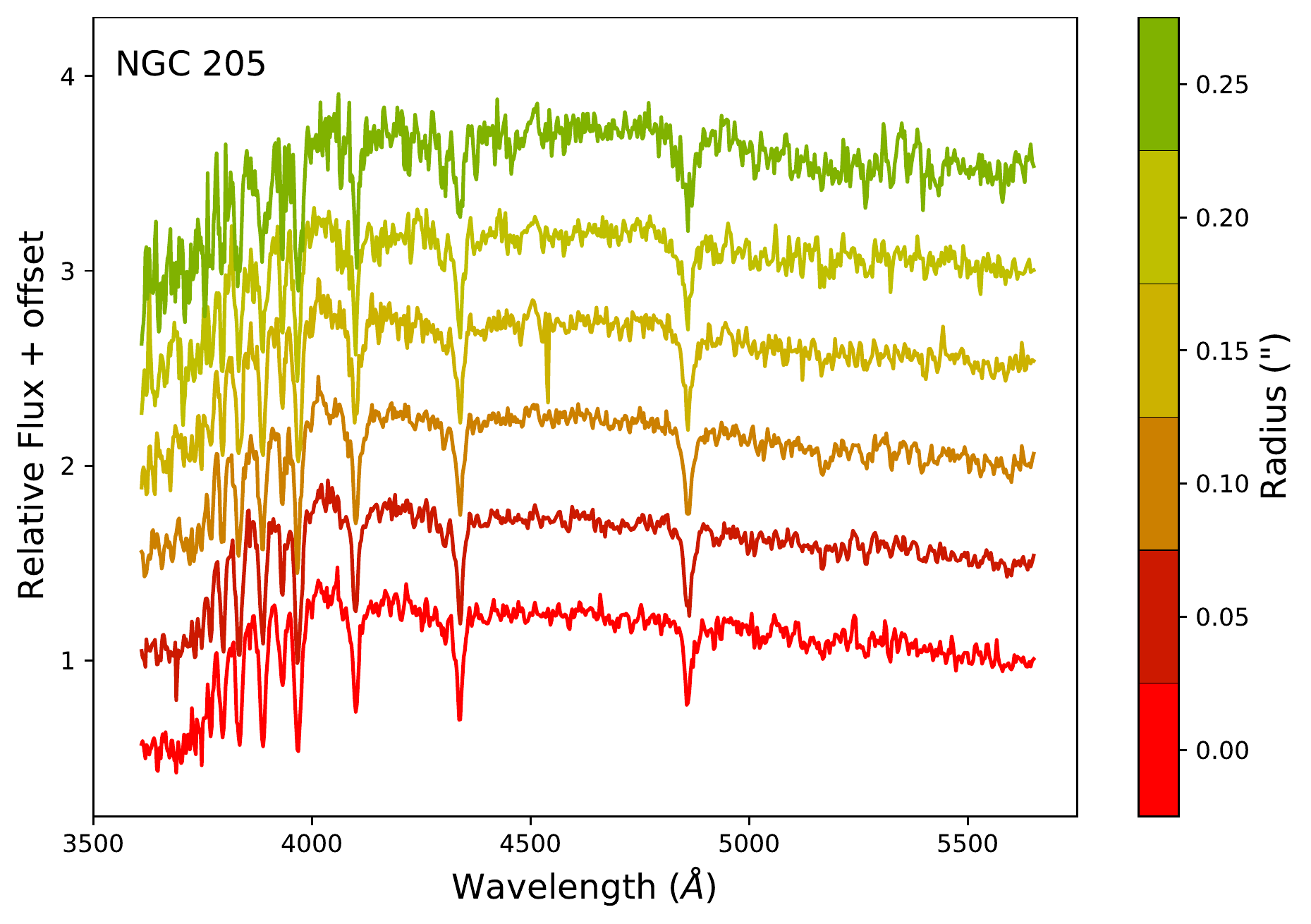}
        \end{minipage}
    \caption{Combined and symmetrically binned spectra at each radius for NGC 5102 (top), NGC 5206 (middle), and NGC 205 (bottom).}
    \label{fig:allspectra}
    \end{flushleft}
\end{figure}

\par
\begin{figure}[htp]
    \centering
    \includegraphics[width=\linewidth]{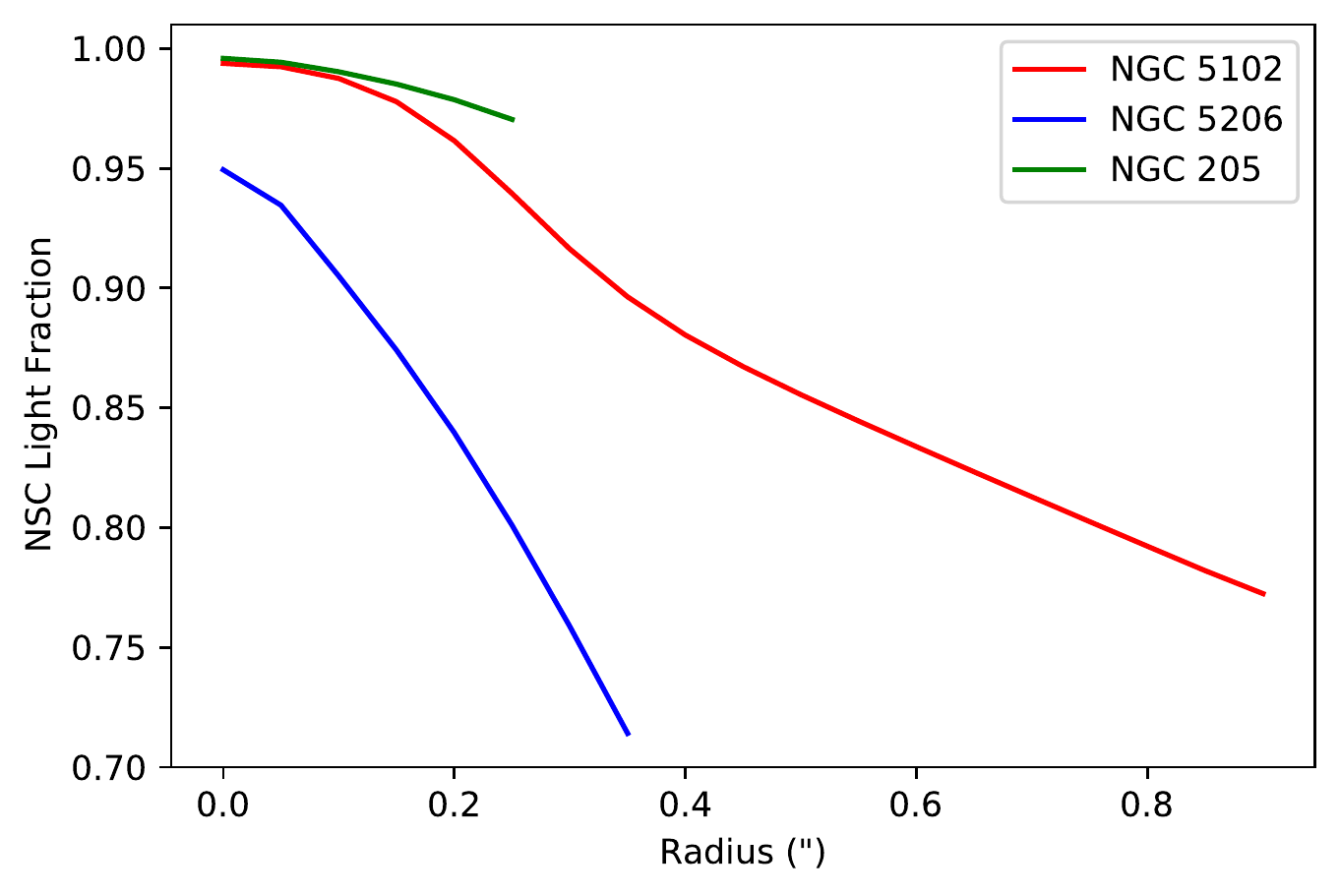}
    \caption{The fraction of light coming from the NSC in each galaxy as a function of radius shows that our STIS data is largely dominated by NSC light out to each galaxy's maximum useful radius.}
    \label{fig:lightfractions}
\end{figure}

 Once the alignment offsets were determined, the images were mean combined with 2$\sigma$ clipping to remove cosmic rays and hot pixels. The standard deviation used for pixel rejection was computed with the addition of two neighboring pixels in the wavelength direction in each image. The 2$\sigma$ threshold was determined by visual inspection of the combined images to ensure artifact removal on the very noisy STIS data with minimal data loss. Figure 1 shows the final combined data for each galaxy after symmetric binning to a median signal-to-noise ($S/N$) threshold of 10. The maximum radii shown for each galaxy corresponds to the largest radii used in this study. Based on the decomposed light profiles for our galaxies \citep[see][]{Nguyen2018}, we note that contamination from galaxy light is minimal at the radii explored here. Figure~\ref{fig:lightfractions} shows the fraction of light coming from the NSC at each radius for all galaxies derived from the S\'{e}rsic profiles given in \citet{Nguyen2018}. To account for the PSF of STIS, the S\'{e}rsic profiles were convolved with a Gaussian PSF (FWHM = 0$\farcs$07) before computing the NSC light fractions.

\subsection{Issues with the Data} \label{sec:issueswithdata}
Here we touch on some data issues which impacted our analysis, including flux calibration and flat fielding issues, as well as error estimation.

\textit{Flux Calibration:}
Careful inspection of the spectra for NGC~5102 (and fits to these spectra) revealed a wavelength dependent flux calibration issue in the images. This issue is most visible in the 0$\farcs$1 - 0$\farcs$15 spectra for NGC 5102 in Figure \ref{fig:allspectra}. No combination of extinction or simple stellar population (SSP) templates provides a decent fit in some regions of the continuum. Because of this issue, the flux calibration was assumed to be incorrect, and we included high order multiplicative polynomials in our stellar population fits (see Section~\ref{sec:cspfits}). This prevented us from examining extinction as a function of radius.

\textit{Flat Fields:}
The STIS flat fields suffer from vignetting near the top of the chip where some of our exposures were focused (for more details see Section~4.1.4 of the STIS Data Handbook\footnote{https://hst-docs.stsci.edu/stisdhb/chapter-4-stis-error-sources/4-1-error-sources-associated-with-pipeline-calibration-steps\#id-4.1ErrorSourcesAssociatedwithPipelineCalibrationSteps-CCDImagingFlatFields}). Because of this, spatial pixels above 1094 were masked during image combination. The center of the galaxies were located in this region of the chip for only one exposure in each set. While larger radii in other images overlapped this masked region, the data loss due to this issue was minimal. 

\textit{Overestimated Errors:}
The median reduced $\chi^2$ ($\chi^2_\nu$) resulting from unregularized composite stellar population (CSP) fits to the combined data at each radius are 0.053, 0.068, and 0.131 for NGC 5102, NGC 5206, and NGC 205, respectively. This suggests the errors in our data are significantly overestimated. The same $\chi^2_\nu$ behavior was seen for data from single exposures as well. An attempt was made to compute the errors manually following the prescription given in the STIS Data Handbook, but the errors did not improve. Therefore, the error in each individual spectrum was scaled to give $\chi^2_\nu$ of 1 for unregularized CSP fits (see Section~\ref{sec:cspfits} for more details).

\section{Methodology} \label{sec:methods}
  
\subsection{Input Simple Stellar Population Models} \label{sec:SSPmodels}
\textit{MILES SSP Models:}
All galaxies in our sample were fit using a grid of MILES simple stellar population (SSP) models\footnote{http://research.iac.es/proyecto/miles/} created by \cite{Vazdekis2010}. This grid covers an age range of [0.063~Gyr, 17.78~Gyr] and total metallicity ([M/H]) range of [-2.32, +0.22]. All models were generated using the MILES empirical stellar spectral library \citep{Sanchez2006}, a set of Padova isochrones \citep{Girardi2000}, and a \citet{Chabrier2003} IMF. Additionally, individual stars used to create these SSP models were selected based on their iron metallicity ([Fe/H]) and follow the abundance pattern of the Milky Way. The MILES SSP grid serves as our nominal set of models, and all fit results presented in this work were obtained using this model grid unless stated otherwise.
\\

\textit{FSPS SSP Models:}
An additional grid of SSP models with varying blue horizontal branch (BHB) fractions was generated using the Flexible Stellar Population Synthesis (FSPS) software\footnote{https://github.com/cconroy20/fsps} \citep[v3.1 for Fortran,][]{ConroyGunnWhite2009, ConroyGunn2010} accessed with the \textsc{python-fsps} bindings of \citet{Foreman-Mackey2014}. This was done to investigate the impact of BHB stars (also referred to as extreme horizontal branch stars) on the stellar population fits (see Section~\ref{sec:BHB}). Because this population of stars is often observed in GCs, cluster inspiral formation could result in the presence of BHB stars within an NSC \citep[e.g.][]{Georgiev2009,Gratton2019}. When present in a stellar population, these stars can strongly affect the Balmer absorption lines that are used in age and metallicity measurements  \citep[e.g.][]{Lee2002,ConroyGunn2010,PercivalSalaris2011,Conroy2018}.
\par The grid of FSPS models used in this work covers a parameter space defined by ages, [M/H]s, and BHB fractions in the ranges [0.0003~Gyr, 14~Gyr], [-1.98, +0.20], and [0.0, 10.0], respectively. The BHB fractions (\textsc{fbhb} in FSPS) represent the fraction of horizontal branch stars that are spread uniformly from an isochrone's red clump out to 10,000 K. Similarly to the MILES SSP models, these models were constructed with a set of Padova isochrones \citep{Girardi2000, Marigo2007, Marigo2008}, a \cite{Chabrier2003} IMF, and the MILES spectral library \citep{Sanchez2006}. 
During construction, the specified BHB fractions are only applied to models older than 3~Gyr.

\subsection{\textit{\textsc{ppxf}} Spectral Fitting Software}
\par The \textsc{ppxf} package\footnote{https://pypi.org/project/ppxf/} \citep[v7.0 for Python,][]{Cappellari2004, Cappellari2017} was used to fit stellar population models to the data for each galaxy. This software implements the Penalized PiXel-Fitting (\textsc{ppxf}) method to conduct full spectrum fitting and can extract stellar/gas kinematics as well as stellar population properties derived from the models. Here we focus solely on the stellar population properties, as the kinematics are not well resolved by our STIS data.

\subsection{Composite Stellar Population Fits} \label{sec:cspfits}
\par Previous spectroscopic measurements show that the star formation histories (SFHs) of NSCs are complex \citep[e.g.][]{Walcher2006,Seth2006,Kacharov2018}.  
For this reason, we performed composite stellar population (CSP) fits at all radii in each galaxy ([0$\farcs$00 - 0$\farcs$90] for NGC~5102, [0$\farcs$00 - 0$\farcs$38] for NGC~5206, and [0$\farcs$00 - 0$\farcs$25] for NGC~205). These CSP fits involved fitting the observed spectrum with a linear combination of SSP models; the weights given to each model were used to measure the mean age, mean metallicity, SFH, and metallicity distribution of the stellar population.

\par We fit all observed spectra over the wavelength range [3610~$\textrm{\AA}$, 5650~$\textrm{\AA}$]. As mentioned in Section~\ref{sec:data}, the data for each galaxy was symmetrically binned\footnote{We also examined bins separately on both sides of the center of each galaxy and found fit results consistent with symmetry.} in the spatial direction to a median $S/N$ threshold of 10. For the binning, we use errors scaled with the median $\sqrt{\chi^2_\nu}$ from initial CSP fits to the unbinned data with original errors. It is also important to note here that the flux calibration issues (see Section~\ref{sec:issueswithdata}) prompted the use of 20$^{th}$-order multiplicative polynomials in all \textsc{ppxf} fits to account for the misshapen continuum (\textsc{mdegree} = 20 in \textsc{ppxf}). A polynomial of degree 20 was determined to provide the best continuum correction without affecting absorption line strengths based on visual inspections of the best-fit polynomials and the decrease of $\chi^2_\nu$ with increasing \textsc{mdegree}.

\par Weights regularization is a property of \textsc{ppxf} that enforces smooth variations in weights of neighboring SSP models \citep{Cappellari2017}. This functionality serves to reduce the degenerate behavior of CSP fit solutions where different sets of SSP models can produce similar $\chi^2_\nu$ values \citep{Kacharov2018}. Because neighboring SSP models share many similarities, a practical motivation exists for enforcing regularization. First-order linear regularization (\textsc{reg\_ord} = 1 in \textsc{ppxf}) was used in this work which ensures the numerical first derivatives of neighboring model weights are within 1/\textsc{regul} of 0. Thus, higher values of \textsc{regul} will result in a smoother distribution of weights.

\par To determine the maximum value of \textsc{regul} in which the solution is still compatible with the data on a 1$\sigma$ level, the value of this regularization parameter was iteratively increased until a $\Delta\chi^2$ threshold was met. First, an unregularized fit of the data was conducted and the error was scaled to give a $\chi^2_\nu$ of 1. This corresponds to a $\chi^2$ of $N$ where $N$ is the number of pixels being fit. Because the variance of the $\chi^2$ distribution is $2N$ ($\sigma = \sqrt{2N}$), \textsc{regul} was increased during successive CSP fits (using the scaled error) until the $\chi^2$ value increased by $\sqrt{2N}$. The solution corresponding to the maximum value of \textsc{regul} is referred to as the regularized solution in this work.

\par After the spatially symmetric binning, the spectra and errors were re-binned to a logarithmic wavelength scale using \textsc{ppxf}'s ``log\_rebin" function. The spectra were also normalized using median fluxes in the wavelength range [5500\AA, 5600\AA]. Using the \textsc{goodpixels} keyword in \textsc{ppxf}, we mask all data points with fluxes $\leq 0$ in all fits. Significant outliers remained in NGC~205, and thus additional masking was performed using the \textsc{ppxf} keyword \textsc{clean} to mask cosmic rays/hot pixels using sigma clipping (see \cite{Cappellari2002} for more details).

\par The STIS spectral resolution has a line spread function full width half max (FWHM) of 5.46\AA (corresponding to $\sigma\sim$150~km/s).  Our model grids are at a higher resolution (FWHM = 2.51\AA) and were log-rebinned with twice the resolution of the data  (\textsc{velscale\_ratio} = 2 in \textsc{ppxf}). 

\par Because the spectra and models are normalized identically, the weights given by CSP fits are light weights (LWs), which provide stellar population parameters based on light contribution. The LWs for fits with the MILES model grid were also converted to mass weights (MWs) via multiplication by the initial mass-to-light (M/L) ratios of the models. MWs give the population parameters based on the initial mass contribution of each SSP.

\par Because this study was not focused on repeating kinematic measurements of the NSCs in these galaxies, the velocity dispersions were manually set with the literature values for NGC 5102, NGC 5206, and NGC 205 shown in Table~\ref{tab1}. The spectral resolution of the STIS data is far too low to resolve the velocity dispersion of any of our galaxies \citep[$\sigma \lesssim 60$~km/s][]{Nguyen2018}.  Therefore, small variations in velocity dispersion have no significant impact on the quality of the fits. The radial velocities were not specified manually in order to obtain the best alignment between the models and data determined by \textsc{ppxf}. This was accomplished by applying an unregularized CSP fit to the central spectrum of each galaxy where the dispersion is fixed. The resulting velocities (which are in rough agreement with literature measurements) remain fixed in all final fits for each galaxy. 

\par In order to aid \textsc{ppxf}'s multiplicative polynomial in correcting issues in the continuum, another unregularized CSP fit was applied to measure extinction for all radii in all galaxies \citep[as in][]{Kacharov2018}. The best-fitting extinction value for each spectrum was then used to correct the model grid. This was accomplished by multiplying the models by the reddening curve of \citet{Calzetti2000} corresponding to the best-fit extinction value. Prior to conducting any final fits, a final unregularized CSP fit was applied to the data using these reddening corrected templates, and the errors were scaled to give a $\chi^2_\nu$ of 1. This scaled error was used in all final fits (both CSP and SSP) along with the extinction corrected templates.

\begin{figure*}[htp]
    \begin{minipage}[t]{0.49\textwidth}
        \includegraphics[width=\linewidth]{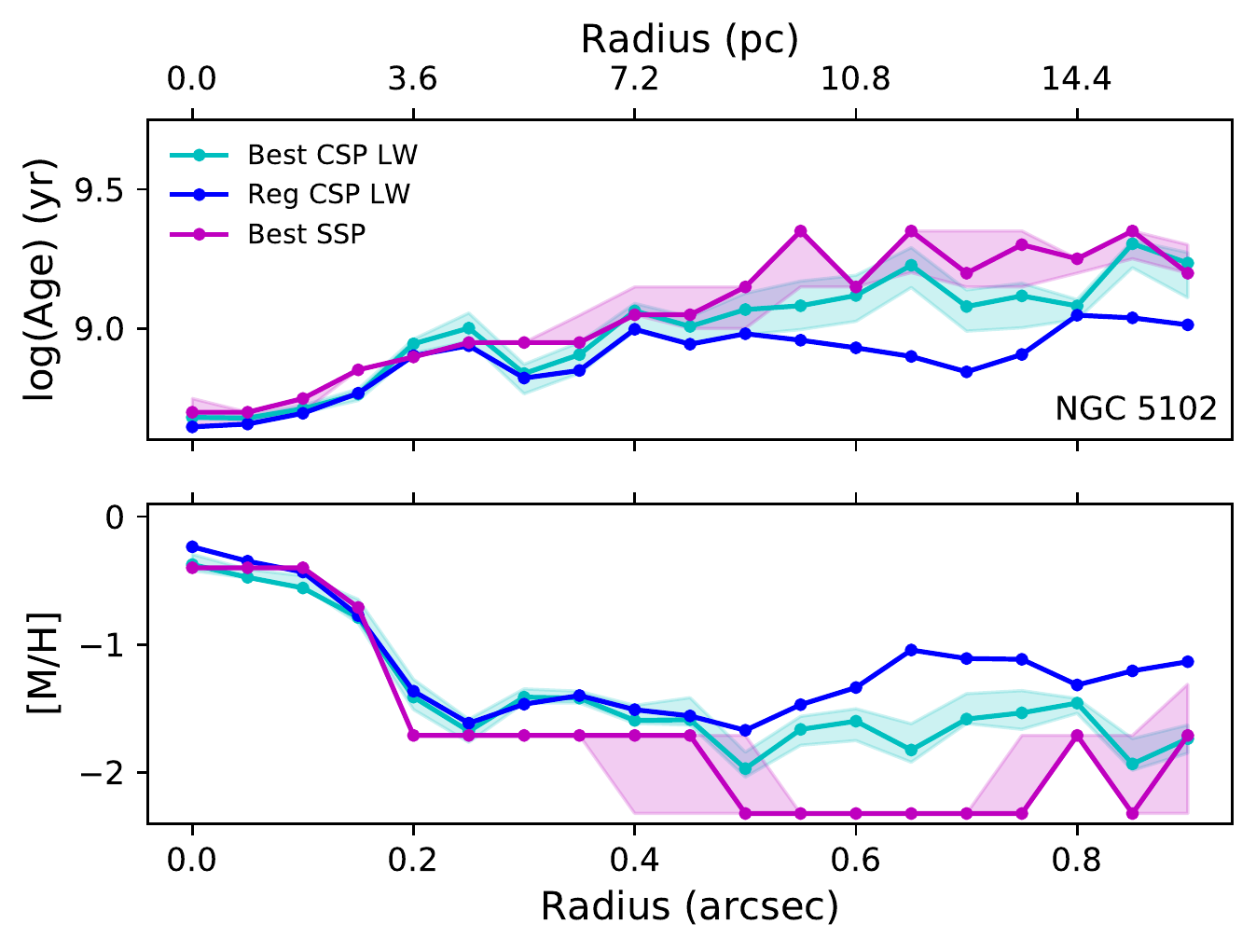}
    \end{minipage} \quad
    \begin{minipage}[t]{0.49\textwidth}                       \includegraphics[width=\linewidth]{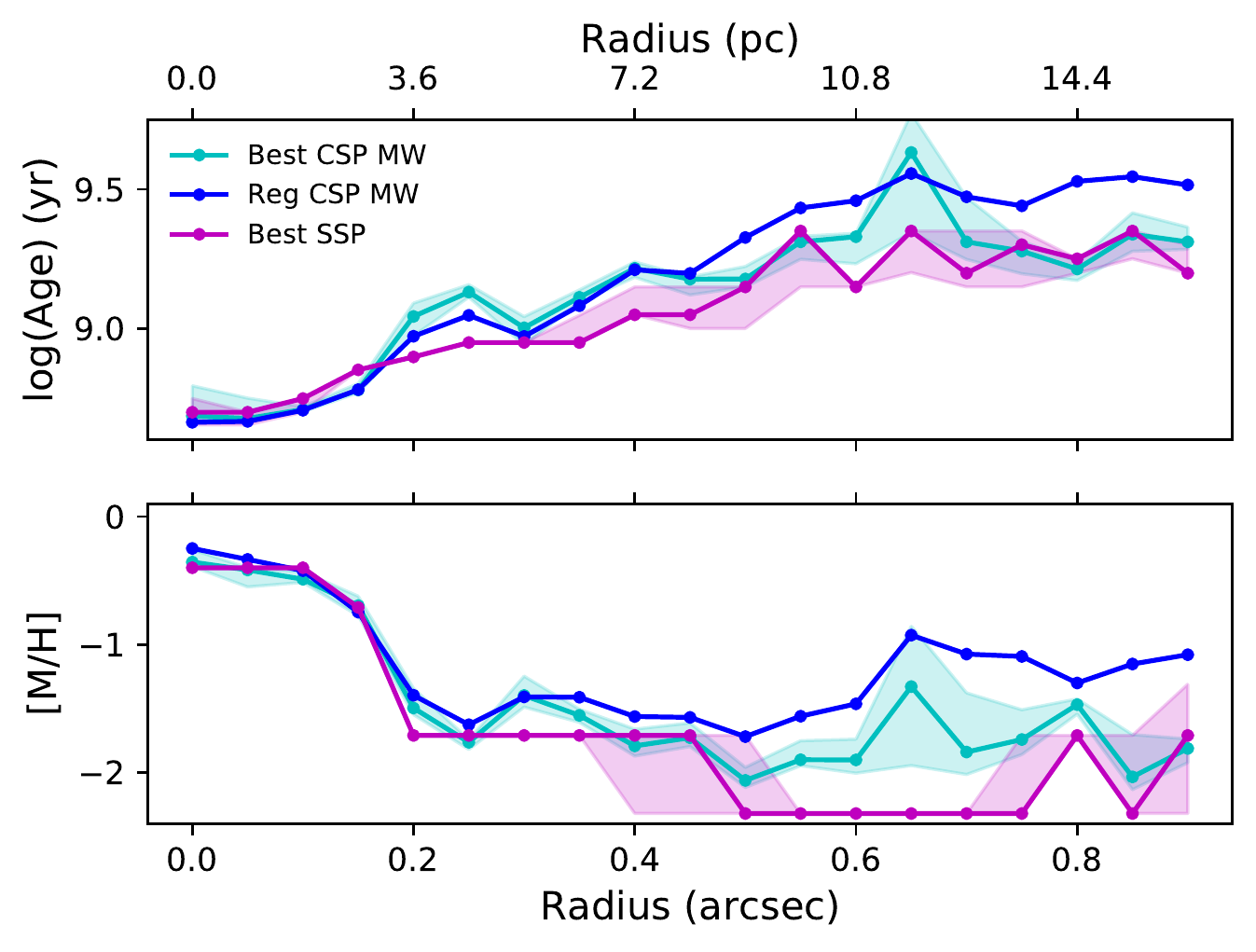}
    \end{minipage}
\caption{The best-fit log(ages) and metallicities as functions of radius for NGC~5102 reveal the presence of both age and metallicity gradients within the NSC. Left: light-weighted unregularized CSP fits with the shaded region giving the 16-th and 84-th percentiles from MC (cyan), light-weighted regularized CSP fit solutions (blue), best SSP fit solutions with the 16-th and 84-th percentiles from MC given by the shaded region (magenta). Right: Same as left but CSP parameters are mass-weighted.}
\label{fig:bestparams}
\end{figure*}

\par In addition to conducting a final regularized CSP fit at each radius, a final unregularized CSP fit was also performed. These fits serve to quantify our errors on the age and metallicity distributions and to investigate the effects of regularization. Uncertainties in the unregularized CSP fits were computed via 20 iterations of a Monte Carlo (MC) simulation. Each MC simulation involved 20 iterations where noise was introduced to the data. This noise was constructed by multiplying the scaled error array by an array of random numbers drawn from a Gaussian distribution with a mean of 0 and standard deviation of 1.

\subsection{Simple Stellar Population Fits} \label{sspfits}
\par SSP fits involve determining the single best-fit model from a grid of SSP models. Essentially, this approach attempts to model the observed spectral energy distribution (SED) as a population of stars born at the same time with identical metallicities \citep{Conroy2013}. We performed these fits as a check on the CSP fit results similar to \citet{Kacharov2018}, which found comparable fit quality between CSP and SSP results for NGC~5102 and NGC~5206. To maintain consistency, the SSP fitting procedure was identical to that of the CSP fits aside from a few required differences. One difference involves the exclusion of weights regularization as this is not applicable in SSP fitting. Also, rather than fitting the data with the full grid of models at once, the data was fit individually with every SSP model. The SSP model resulting in the lowest $\chi^2_\nu$ was selected as the best-fitting SSP model. We computed uncertainties on the best SSP parameters using MC simulations similar to those used in our CSP fits (see Section~\ref{sec:cspfits}).

\section{NGC~5102 Results} \label{sec:results}
\subsection{CSP Fits} \label{sec:cspresults}

\par The mean population parameters derived from our unregularized CSP fits can be seen in Figure~\ref{fig:bestparams} labeled as ``Best CSP LW" and ``Best CSP MW" for light-weighted and mass-weighted quantities, respectively. The corresponding shaded regions give the 16-th and 84-th percentiles of the parameters resulting from the unregularized MC simulations described above. These results reveal the presence of substantial age and metallicity gradients within the NSC of NGC~5102 where the stellar populations are young and metal-rich in the very center but become older and more metal-poor at larger radii.

\par The regularized CSP fit solutions are also displayed in Figure~\ref{fig:bestparams} labeled ``Reg CSP LW" and ``Reg CSP MW". While these results agree well with our unregularized CSP fits at lower radii, it is worth noting here that the regularized solutions are slightly biased towards the median age and metallicity of the model grid at large radii due to the lower $S/N$ and larger regularization parameters. All results presented in Figure~\ref{fig:bestparams} are listed in Table~\ref{tab:NGC5102_Results_ALL}. 

\begin{figure*}[t]
    \begin{minipage}[t]{0.49\textwidth}
        \includegraphics[width=\linewidth]{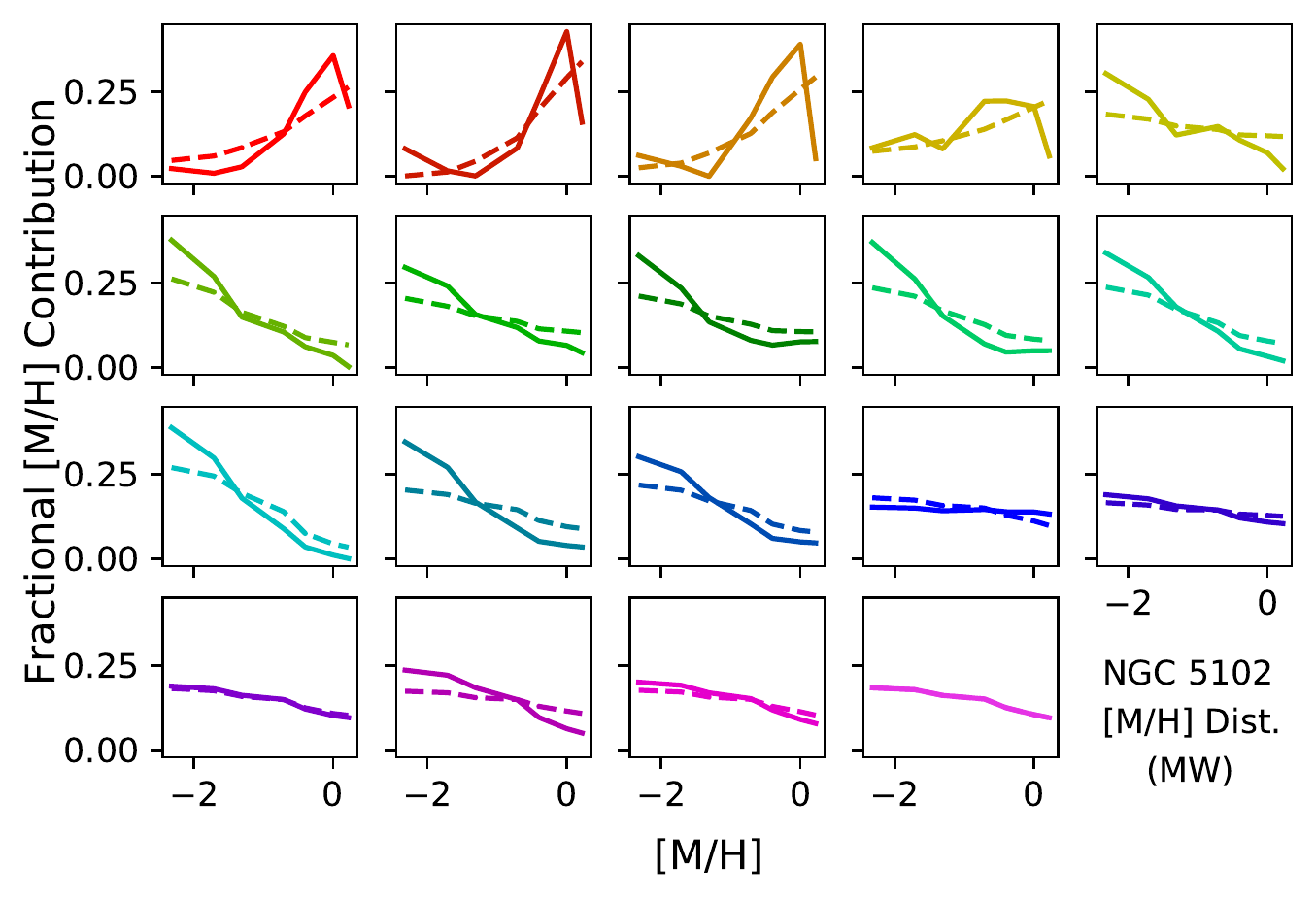}
    \end{minipage} \quad
    \begin{minipage}[t]{0.49\textwidth}                             
        \includegraphics[width=\linewidth]{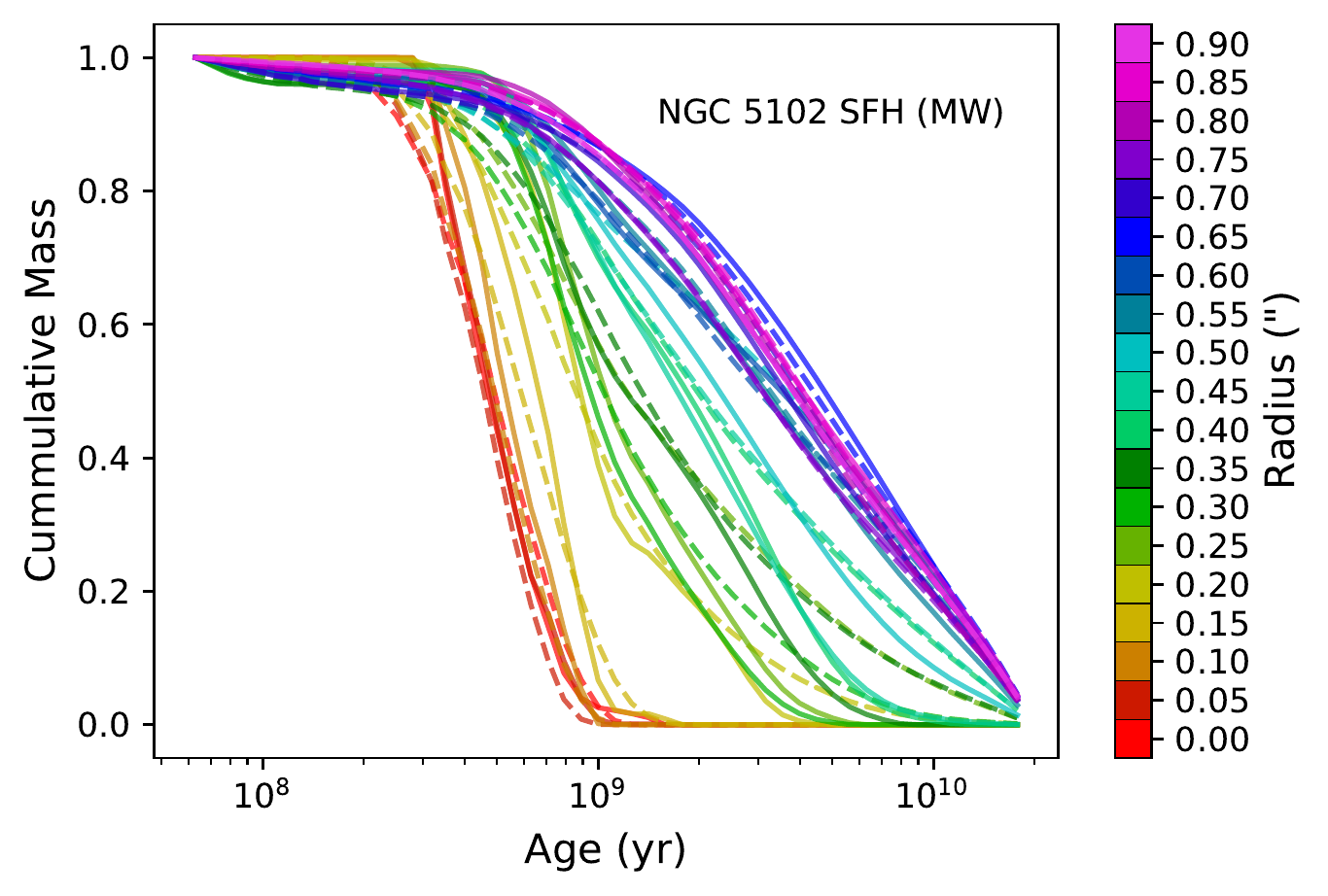}
    \end{minipage}
\caption{The mass-weighted SFHs at different radii within NGC~5102 (right) show the NSC gets older with increasing radius. Note that the y-axis shows the fraction of mass formed at ages older than the x-axis, and the cumulative masses are in initial (not present day) masses. The mass-weighted metallicity distributions (left) show the declining contribution of metal-rich SSPs with increasing radius. These results use the regularized solutions, and thus are biased to similar values at low $S/N$. The dashed lines in both plots show results from additional regularized fits using the same level of regularization as the outermost (and lowest $S/N$) spectrum for all radii. These excessively smoothed results demonstrate that the observed trends in age and metallicity are not due to regularization.}
\label{fig:SFH}
\end{figure*}

\par Figure~\ref{fig:SFH} displays the mass-weighted metallicity distributions (left) and SFHs (right) derived from the model weights given by the regularized CSP fits at each radius. The solid lines give the results from the standard regularized fits described in Section~\ref{sec:cspfits}, while the dashed lines show the results of regularized fits using the largest regularization parameter (corresponding to the 0$\farcs$9 spectrum) for all radii. Using increased regularization at all radii effectively increases the smoothing of the model weights for each solution. Despite this, the age and metallicity trends remain apparent in the metallicity distributions and SFHs (see Figure~\ref{fig:bestparams}). We show in Section~\ref{sec:robustness} that the gradients are also not a product of the decreasing $S/N$.
More specifically, we find that the stellar populations closer to the center of the NGC~5102 NSC exhibit a shorter and more recent period of star formation than the populations at larger radii.
The metallicity distributions of the inner populations also peak around solar metallicity and shift towards lower metallicities at radii $\geq$0$\farcs$2.

\subsection{SSP Fits} \label{sec:sspresults}
\par The best-fit population parameters from SSP fits are shown in Figure~\ref{fig:bestparams} as well with the shaded region indicating the 16-th and 84-th percentiles from MC simulations (labeled ``Best SSP"). Because light- and mass-weighted parameters don't apply for SSP fits, the SSP data shown in both plots are identical. We find that the SSP fit results agree very well with the trends in our CSP solutions. Specifically, the central spectrum is best-fit by an SSP of 0.5~Gyr and [M/H] = -0.40, while the spectra at radii between 0$\farcs$3 and 0$\farcs$9 are on average best-fit by a 1.5~Gyr and [M/H] = -2.32 model.  We discuss whether the NGC~5102 NSC is well fit by a model combining just these two SSPs in Section~\ref{sec:sersic}.  

\input{NGC5102_TABLE}
\vspace{-24pt}

\subsection{Robustness of Gradients} \label{sec:robustness}
\par Because the $S/N$ of our data decreases with radius (from $\sim$60 at the center to $\sim$20 at 0$\farcs$90 with scaled errors), we developed a test to ensure the observed age and metallicity trends were not created simply by the decrease in $S/N$. This test involved taking the best-fit model from the regularized solution at the center (LW age $\sim$400~Myr, LW [M/H] $\sim$-0.3) and degrading the $S/N$ computed with the scaled errors at each radius to match the $S/N$ of the data. Unregularized CSP fits were then performed for this simulated data to see if we would recover similar trends in age and metallicity. 

\par
\begin{figure}[!t]
    \centering
    \includegraphics[width=\linewidth]{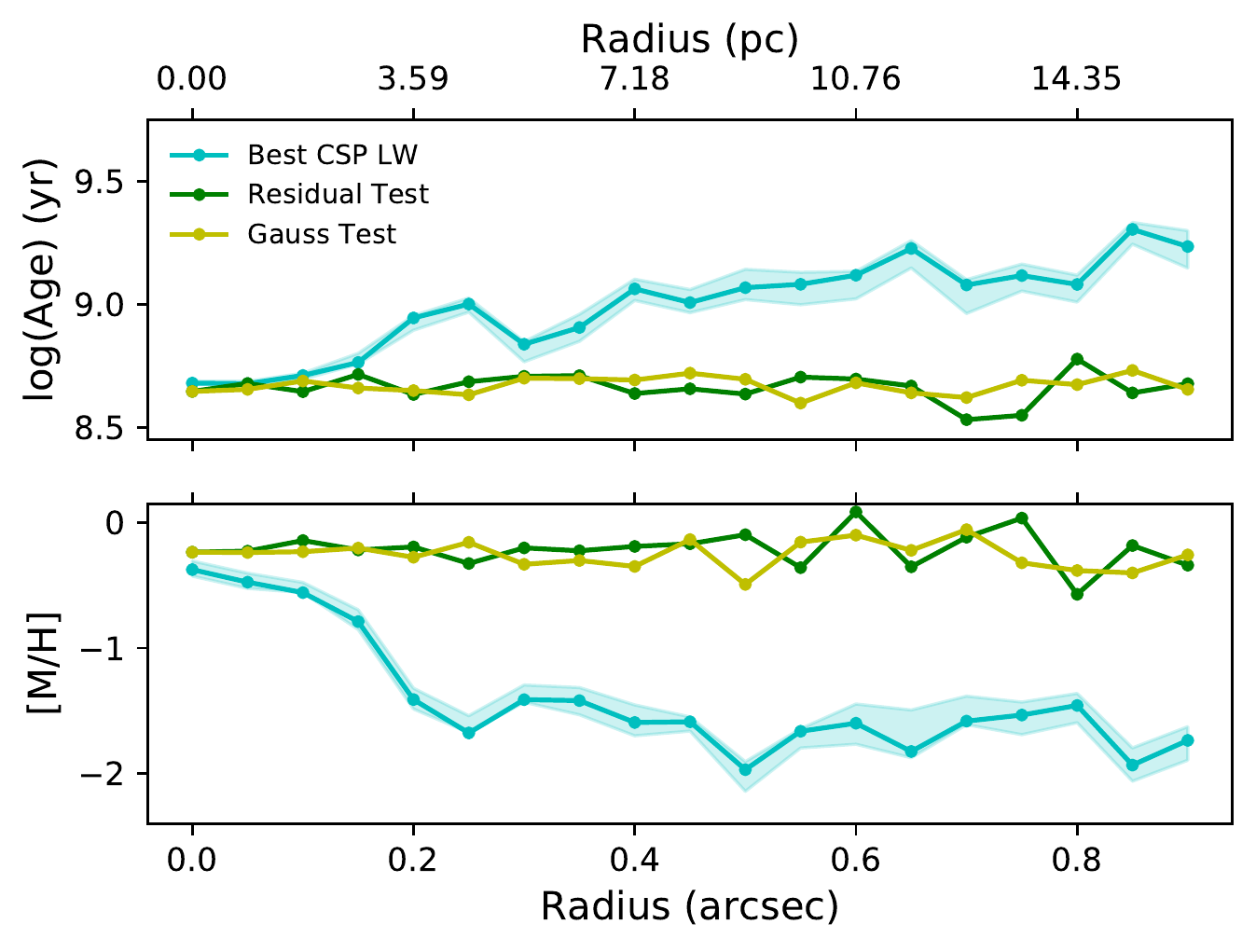}
    \caption{Light-weighted mean log(ages) and metallicities as functions of radius for NGC~5102 resulting from unregularized CSP fits to the best-fit regularized CSP solution at the center of the galaxy that was degraded with Gaussian noise (Gauss Test) and scrambled residuals (Residual Test) to match the $S/N$ of our data at each radius (see Section~\ref{sec:robustness} for more details). These results are compared to the light-weighted CSP fit results from the data (Best CSP LW, same as in Figure~\ref{fig:bestparams}), and reveal that the observed gradients are not an effect of decreasing $S/N$.}
    \label{fig:trendtest}
\end{figure}

 The $S/N$ was degraded in two different ways resulting in two different (but similar) tests. The first method involved adding Gaussian noise to the model with a standard deviation given as $\sigma = \sqrt{|\sigma_{\textrm{res},i}^2 - \sigma_{\textrm{res},0}^2|}$ where $\sigma_{\textrm{res},0}$ is the standard deviation of the residuals from the regularized CSP fit at the center and $\sigma_{\textrm{res},i}$ is the same but for the regularized CSP fit at the $i$-th radial bin. The second method involved adding the scrambled residuals from regularized CSP fits at each radius to investigate the effects of outlier pixels as well. Figure~\ref{fig:trendtest} shows the results of these tests where fits to the model data degraded with Gaussian noise are labeled ``Gauss Test" and fits to the model data degraded with scrambled residuals are labeled ``Residual Test." The results are shown along with the best CSP fit parameters to the real data (same as in Figure~\ref{fig:bestparams}). 

\par The flat profiles seen in fit parameters resulting from the simulated data sets indicate that the $S/N$ trend present in our data is not contributing to the observed trends in population parameters. As expected, the best-fit parameters resulting from the ``Residual Test" show slightly more variation than those coming from the ``Gauss Test." This suggests that outliers in the data may slightly effect our parameter measurements, but not on a scale that could explain the observed gradients.

\par To further verify the trends in age and metallicity, unregularized CSP fits were performed using an additional set of MILES SSP models constructed with BaSTI isochrones \citep{Pietrinferni2004} (again using a \citet{Chabrier2003} IMF) and FSPS models without increased BHB fractions. When compared to our previous results, the MILES-BaSTI model grid provided marginally younger ages and higher metallicities at most radii in the galaxy, which is consistent with the age-metallicity degeneracy of SEDs. The FSPS models gave slightly older ages with higher metallicities than the MILES-BaSTI models. The FSPS models also provide the youngest age and highest metallicity for the center of the galaxy. We note that these differences in age and metallicity were minor, and overall, both model sets provided age and metallicity gradients that are consistent with the ones shown in Figure~\ref{fig:bestparams}. 

\subsection{Comparisons to Previous Work} \label{sec:comparisons}

\subsubsection{Integrated Light Fits}

\par In this section, we compare our stellar population synthesis results to previous results obtained from integrated light spectra for NGC~5102. \cite{Kacharov2018} performed a regularized CSP fit (following an almost identical procedure as described in  Section~\ref{sec:cspfits}) to long-slit X-Shooter spectroscopic data that was binned together within a 1$''$ aperture. To perform a comparable fit, our data was binned together out to the maximum useful radius of 0$\farcs$9 and a regularized CSP fit was applied. 

\par
\begin{figure}[!t]
    \centering
    \includegraphics[width=\linewidth]{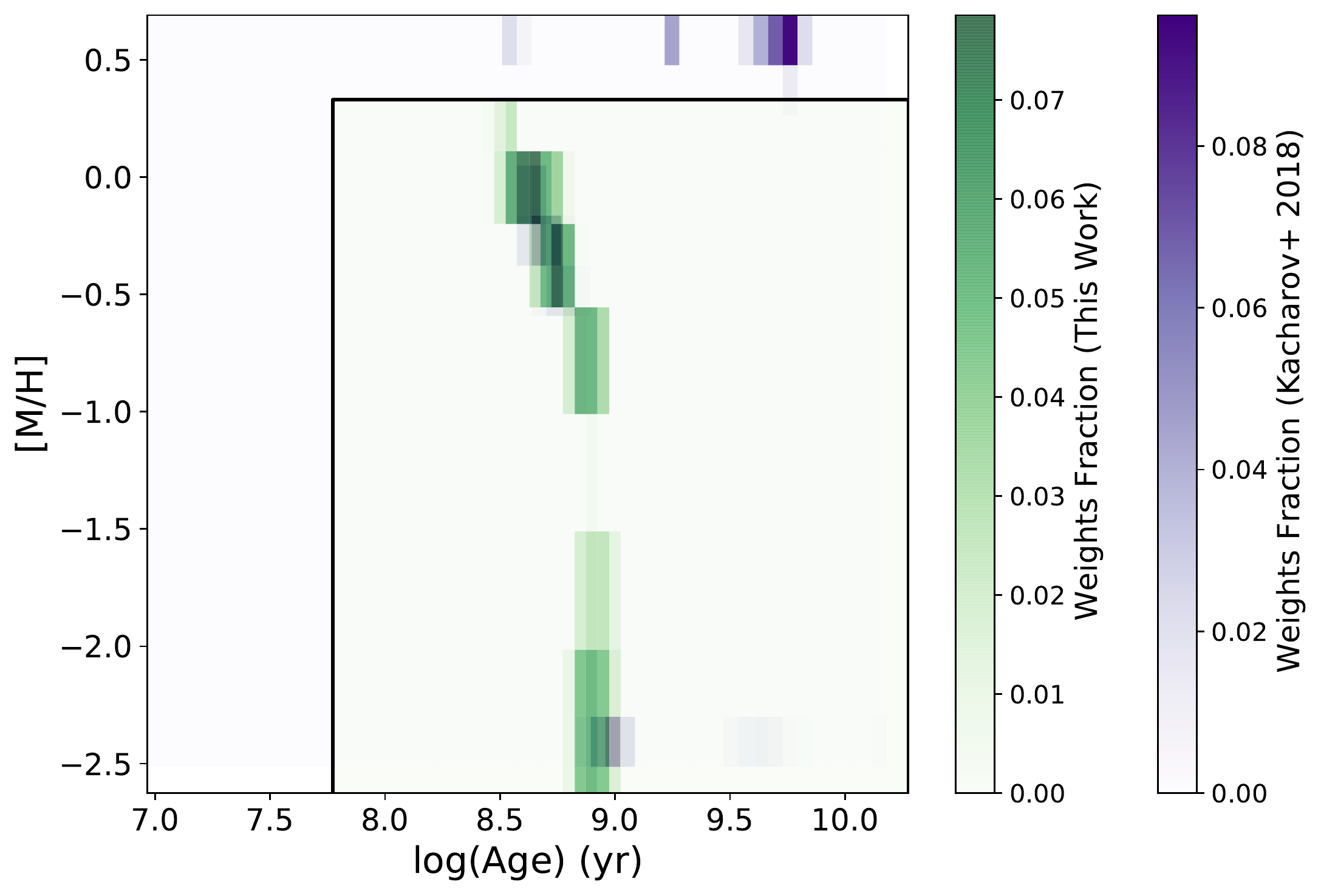}
    \caption{Normalized mass weights from integrated light spectral synthesis fits to our STIS data (green) and the XShooter data presented in \cite{Kacharov2018} (purple).  Both are  regularized CSP fits. The solid black line gives the boundary of the MILES model grid used here. Although the y-axis is labeled [M/H], we note that the metallicities of the \textsc{pegase-hr} models used by \citet{Kacharov2018} are based on [Fe/H] with scaled solar abundance ratios. However, the differences between these metallicity definitions are smaller than the resolution of the model grids.}
    \label{fig:integratedlight}
\end{figure}

 Our results can be seen overlaid with the results of \cite{Kacharov2018} in Figure~\ref{fig:integratedlight} which shows the distribution of mass weights over the model grids used in the fits. Although the data of \cite{Kacharov2018} was much higher resolution (FWHM = 0.2\AA/pix), and a different grid of SSP models \citep[\textsc{pegase-hr};][]{LeBorgne2004} was used, our regularized CSP fit recovered an almost identical distribution of SSPs (aside from the SSPs with parameters outside of our model grid). 

\par We do not recover the highest metallicity components of the population seen in the \cite{Kacharov2018} results. Given that these populations are both old ($\sim$ 5~Gyr) and very metal-rich ([M/H] $\sim$ +0.55), the presence of these SSPs in the NGC~5102 NSC seem unlikely. If an old, metal-rich component was truly present in the population, it would likely still be visible in our results with a similar age and the maximum metallicity of our model grid. This suggests that these old metal-rich SSPs may be a result of errors in the data, fitting procedure, or \textsc{pegase-hr} models. Aside from this minor disagreement, our results are consistent with the findings of \cite{Kacharov2018}. 

\begin{figure*}[t]
    \begin{minipage}[t]{0.49\textwidth}
        \includegraphics[width=\linewidth]{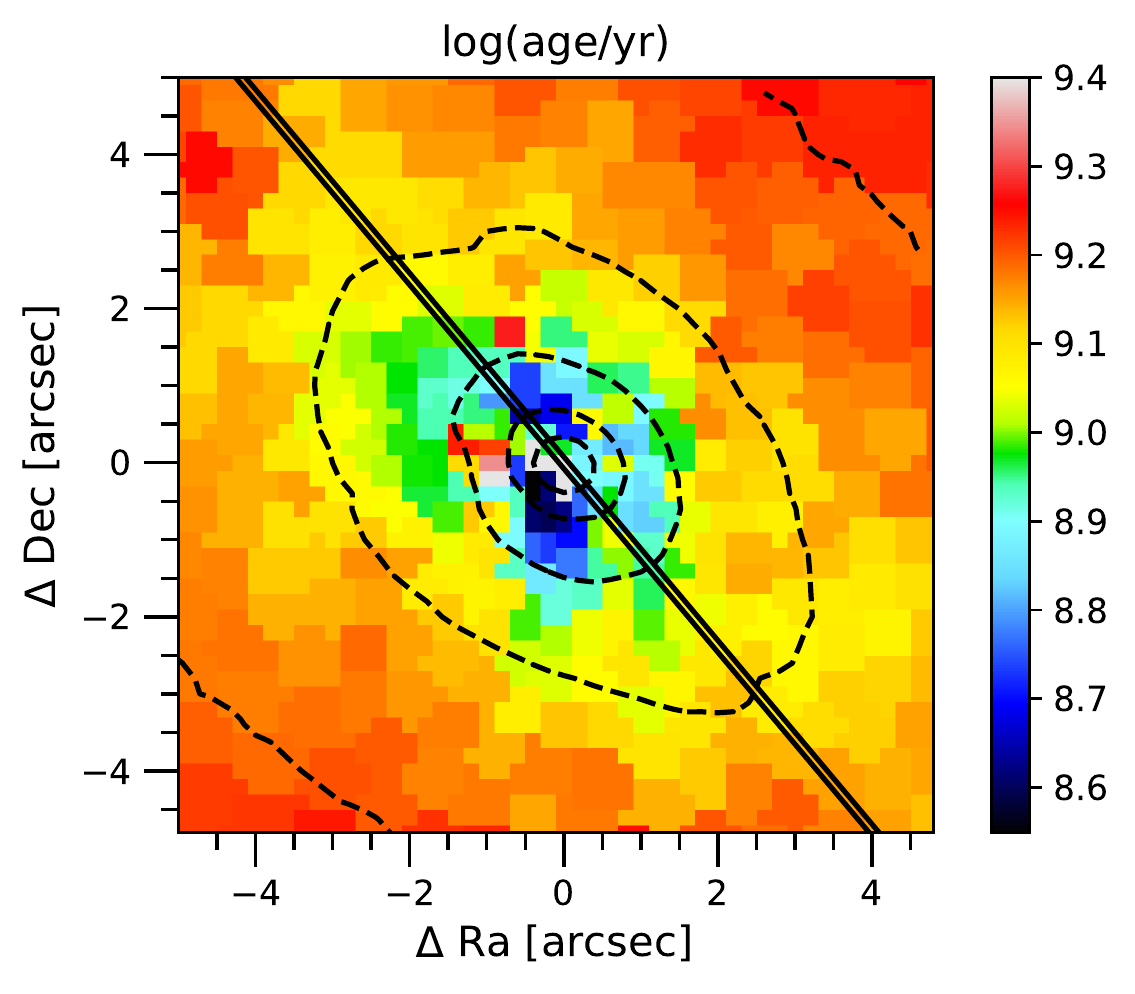}
    \end{minipage} \quad
    \begin{minipage}[t]{0.49\textwidth}                             \includegraphics[width=\linewidth]{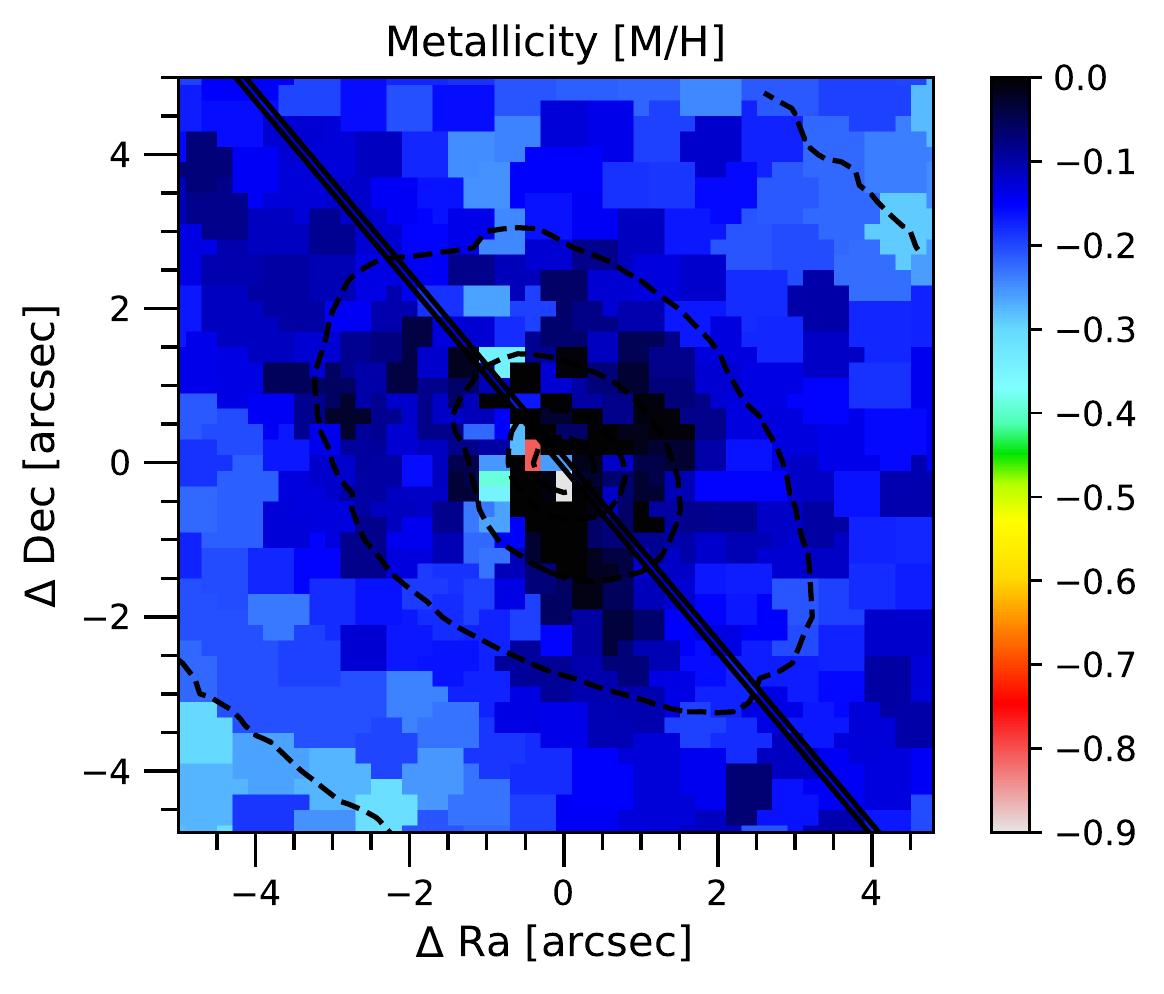}
    \end{minipage}
\caption{Regularized mass-weighted log(age) and metallicity maps for NGC~5102 from MUSE spectral synthesis fits in \cite{Mitzkus2017}. 
Dashed black lines show the isophotes of the galaxy's integrated light while the solid black lines show the approximate location of the STIS slit from this work. Despite the lower resolution, these maps show a drop in metallicity and increase in age just outside the center similar to what we see in our measurements.}
\label{fig:maps}
\end{figure*}

\begin{figure}[!t]
    \begin{flushright}
        \begin{minipage}[t]{\linewidth}
        \end{minipage}
        \hfill
        \begin{minipage}[t]{\linewidth}
            \includegraphics[width=\linewidth]{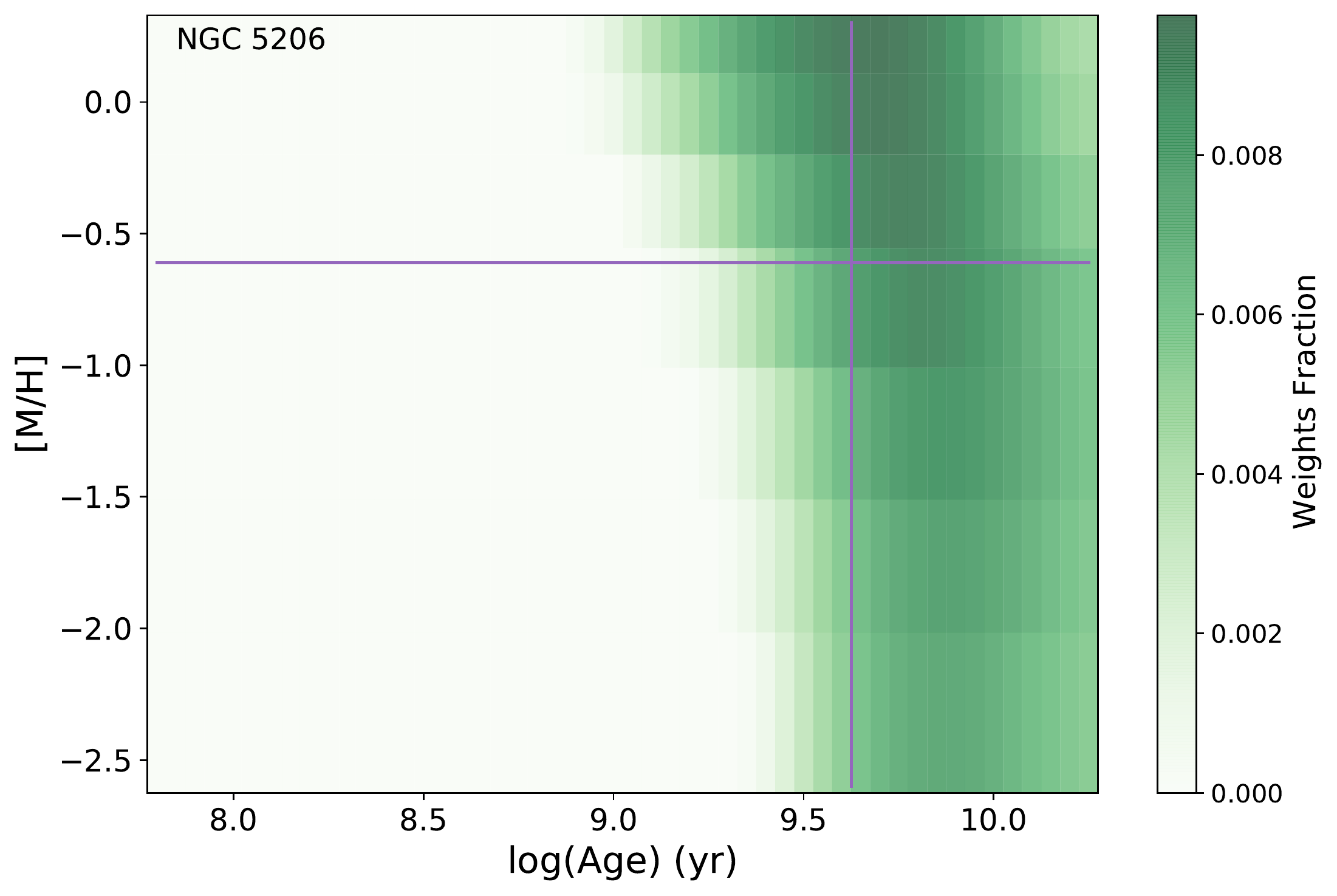}
        \end{minipage}
        \hfill
        \begin{minipage}[t]{\linewidth}
            \includegraphics[width=\linewidth]{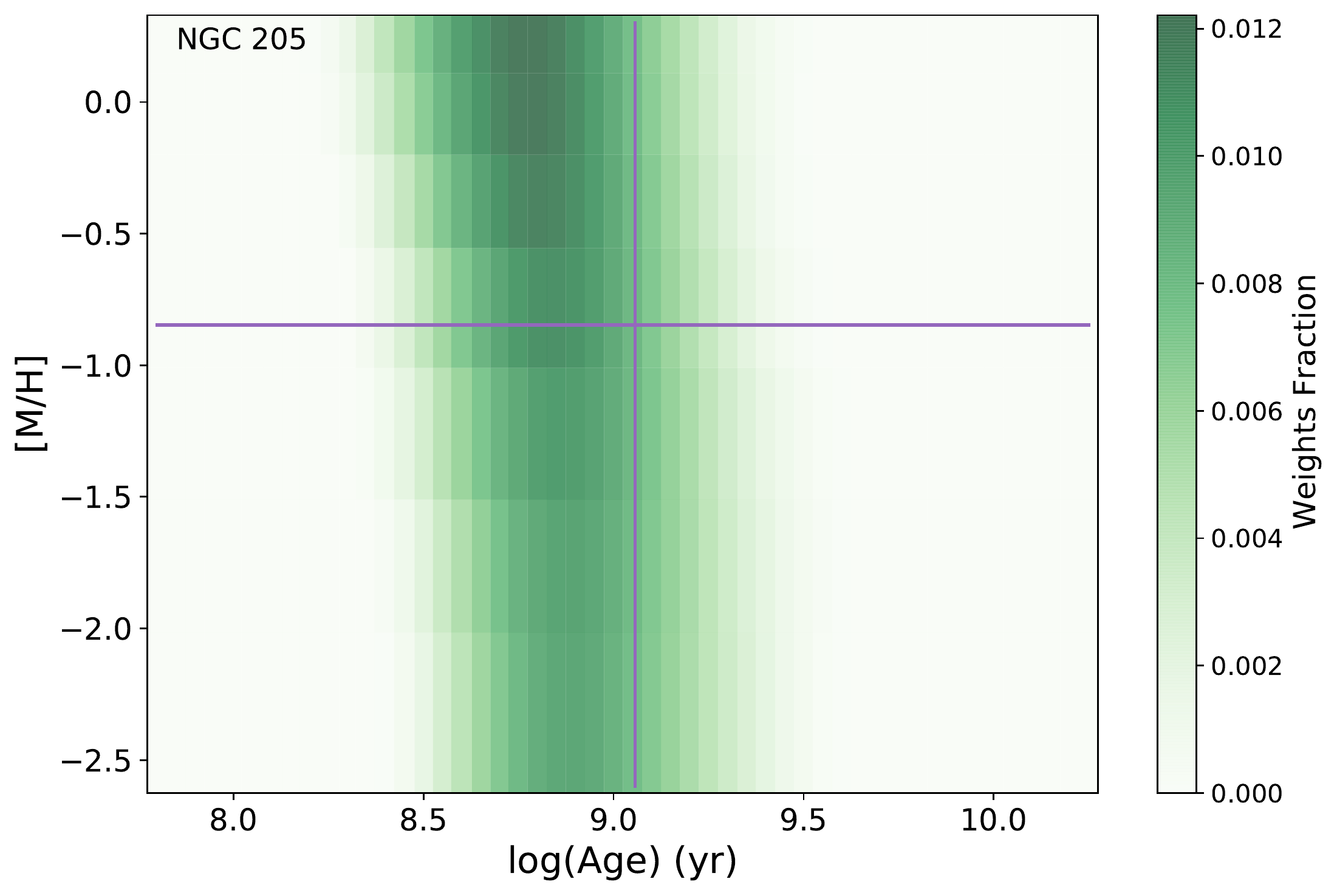}
        \end{minipage}
    \caption{Normalized light weights from  regularized CSP fits to the integrated light of NGC~5206 (top; regul=163) and NGC~205 (bottom; regul=212). Best-fit log(ages) and metallicities for each galaxy resulting from unregularized CSP fits are indicated with the purple lines. The integrated spectra include light out to the maximum useful radius of each galaxy (see Section~\ref{sec:data}).}
    \label{fig:205and5206results}
    \end{flushright}
\end{figure}

\subsubsection{Age \& Metallicity Maps}

\par The age and metallicity maps for NGC~5102 generated by \cite{Mitzkus2017} can be seen in Figure~\ref{fig:maps}. These maps were generated by fitting \textsc{miuscat} (\cite{Vazdekis2012}) SSP models with \textsc{ppxf} to integral-field spectroscopic data obtained with VLT's MUSE instrument. The MUSE data covers the wavelength range [4750~\AA - 9340~\AA], and the seeing was $\sim0\farcs6$ at the time of observation. The very center of the age map reveals the presence of both old and young populations within the maximum radius investigated here. The ages found in this region are also in rough agreement with the ages found in this study. Additionally, the center of the metallicity map reveals metallicities consistent with our results. Most metallicity values measured within our maximum radius fall between +0.0 and -0.3 with a few locations having lower metallicities around [M/H] = -0.9. This provides additional evidence for the presence of an older, metal-poor population close to the center of the NSC. Minor inconsistencies between these results are likely due to the increased spatial resolution of our STIS data.  We also note that we are only analyzing light through a 1$''$ wide slit along the major axis of the NSC.

\section{NGC~5206 \& NGC~205 Results} \label{sec:otherresults}

\par From regularized CSP fits to the integrated light of the NSCs (Figure~\ref{fig:205and5206results} and Tables~\ref{tab:NGC5206_Results_ALL}~\&~\ref{tab:NGC205_Results_ALL}), we find that NGC~5206's NSC is the oldest of the clusters here (LW age $\sim$5.8~Gyr), with LW metallicity of $\sim$ -0.72. The best-fit LW age ($\sim$4.2~Gyr) and metallicity ($\sim$-0.61) resulting from unregularized CSP fits for NGC~5206 (also shown in Figure~\ref{fig:205and5206results} and Table~\ref{tab:NGC5206_Results_ALL}) are consistent with the regularized solutions. These integrated results are in rough agreement with the LW age of 2.91~Gyr and metallicity of -0.32 given by regularized fits in \citet{Kacharov2018}. SSP fits to the integrated light of NGC~5206 provide an age of $\sim$5.0~Gyr and [M/H] of $\sim$-0.71. These SSP results are consistent with our CSP results but disagree with the SSP results of \citet{Kacharov2018} (age = 2.05~Gyr, [Fe/H] = -0.16); although these are likely similar fits based on the expected age-metallicity degeneracy. 

\par For the NGC~205 NSC, we find a LW age $\sim$0.73~Gyr and a LW metallicity of $\sim$-0.78 from  regularized CSP fits, while the unregularized fits provide a LW age of $\sim$1.1~Gyr and metallicity of $\sim$-0.84 (Figure~\ref{fig:205and5206results} and Table~\ref{tab:NGC205_Results_ALL}). This is the first full spectrum stellar population fit to NGC~205 presented in the literature. Previous information on the stellar population is available from the index measurements of \citet{Sharina2006} and photometry near the center from \citet{Butler2005} and \citet{Monaco2009}. Although these studies include light from larger radii in the galaxy, they generally find metallicities $\gtrsim$ -0.9 for NGC 205, which is consistent with our results. \citet{Monaco2009} also find evidence for a population of stars that formed $<$ 1~Gyr ago in the center of NGC~205 similar to our integrated age estimates. We note in both galaxies that the errors on our age and metallicity measurements are large due to the low S/N and a significant age-metallicity degeneracy.

\par From the spatially resolved fits to the NGC~5206 NSC, we find a metallicity gradient that is similar to the one observed in NGC~5102 (see Figure~\ref{fig:bestparams5206}). However, the significance of this result is much lower than for NGC~5102 due to the lower $S/N$ data. The resolved fits for NGC~205's NSC, shown in Figure~\ref{fig:bestparams205}, do not reveal any notable age or metallicity gradients. Because these fits do not provide compelling evidence of stellar population gradients in these NSCs, we focus our discussion on the resolved properties of NSCs to NGC~5102. We direct the reader to Appendices~\ref{app:5206} and \ref{app:205} for more details on the resolved fits for the NSCs of NGC~5205 and NGC~205.

\section{Discussion} \label{sec:disscusion}

\subsection{Formation Mechanisms}

\par The age and metallicity gradients found within the NSC of NGC~5102 offer insight to the formation history of the cluster (Figure~\ref{fig:bestparams}). The presence of a young (LW age $\sim$400~Myr) and relatively metal-rich (LW [M/H] $\sim$-0.4) population at the very center suggests a period of \textit{in situ} star formation in this region. We note that this population could result from the accretion of young star clusters that formed near the galaxy center. However, the high central concentration of this population (shown in Figure~\ref{fig:bestparams}) further argues for \textit{in situ} formation. The older (LW age $\sim$1~Gyr) and metal-poor (LW [M/H]$\sim$-1.6) populations present at radii between 0$\farcs$3 and 0$\farcs$9 are consistent with the cluster inspiral formation mechanism. This metallicity is much lower than both the very central regions of the NSC in our data, and the dominant population for the central regions of the galaxy from \cite{Mitzkus2017} (Figure~\ref{fig:maps}), which has an average [M/H] of roughly -0.2 (but with some lower metallicity spaxels at similar radii to where we find low metallicities). These low metallicities suggest that these populations likely formed in GCs at larger radii that migrated into the NSC. The ages found for these older populations are rather young when compared to the ages expected for GCs. However, age estimates given via stellar population synthesis are less constrained than metallicity measurements due to the spectral similarity between older SSP models. Furthermore, the mass-weighted ages are even older at larger radii because of higher M/L ratios for the older SSP models.  These older ages are more consistent with cluster inspiral formation scenario.  We explore below in Section~\ref{sec:BHB} whether the young best-fit ages we find in the outer parts of the NGC~5102 NSC could be due to BHB stars.  

\begin{figure}[t]
    \centering
    \includegraphics[width=\linewidth]{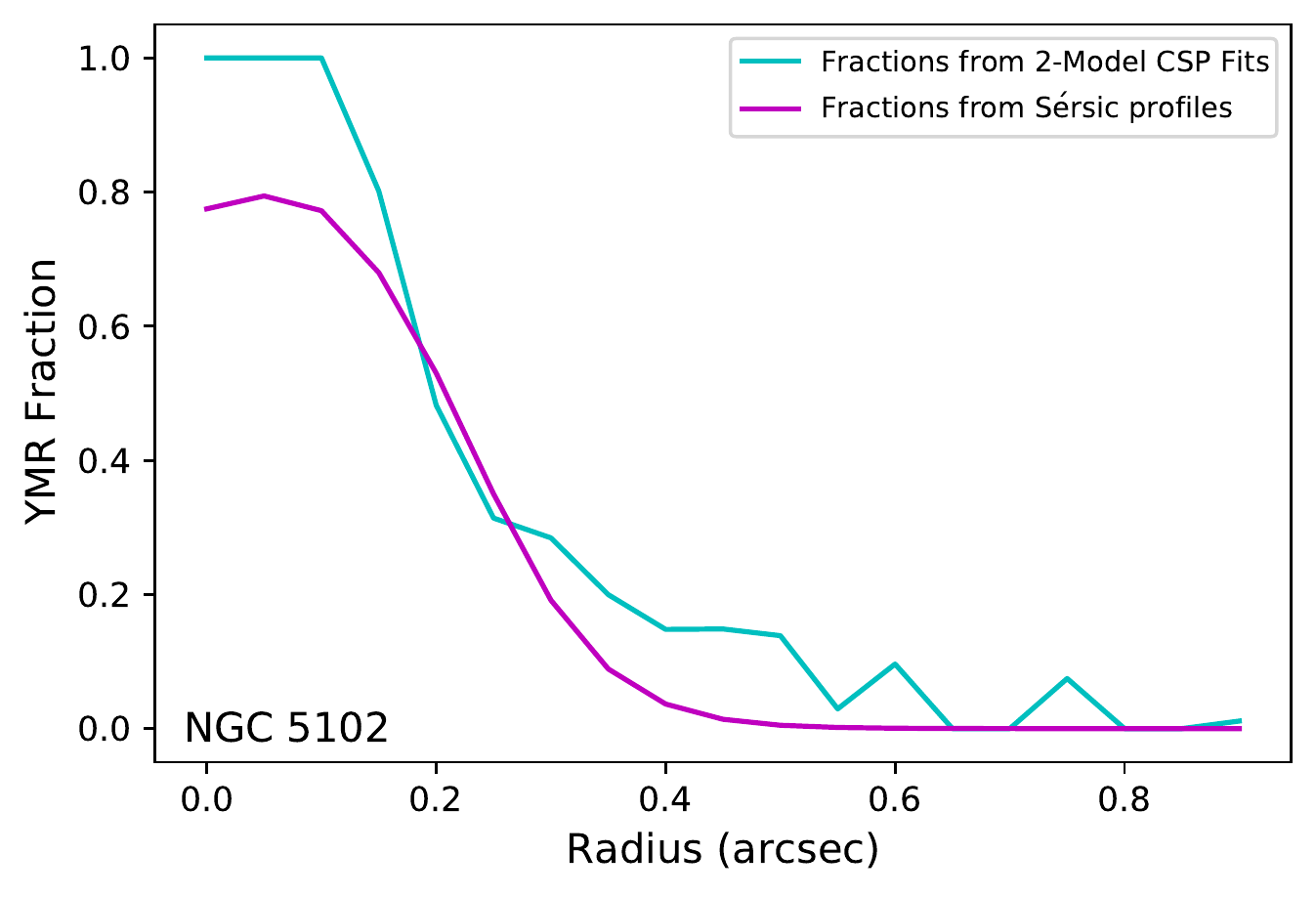}
    \caption{Fractional contributions of a central YMR population given from CSP fits to the data using only two MILES SSP models (cyan) and from the PSF convolved S\'{e}rsic profiles of \cite{Nguyen2018} (magenta).}
    \label{fig:lightprofiles}
\end{figure}

\subsection{Relation to Surface Brightness Profile} \label{sec:sersic}

\par In our results, the transition between the young, metal-rich population at the center to the older, metal-poor population occurs at a radius of roughly 0$\farcs$20. Additionally, when performing S\'{e}rsic profile fits to the surface brightness (SB) profile of NGC~5102, \cite{Nguyen2018} find two distinct NSC components that intersect around this same radius with the following parameters: Inner - n = 0.8, R$_\textrm{e}$ = 0$\farcs$1; Outer - n = 3.1, R$_\textrm{e}$ = 2$\farcs$0. 

\par To investigate the relation between these results, a test was conducted where unregularized CSP fits using only two MILES SSP models were applied to our data. The two models used were the best-fit SSP at the center (0.5~Gyr, -0.40 [M/H]) and the closest SSP corresponding to the average age and metallicity of the best-fit SSPs in the radial range [0$\farcs$3, 0$\farcs$9] (1.5~Gyr, -2.32 [M/H]). The fractional contribution of the young, metal-rich (YMR) SSP to the CSP fit at each radius was then compared to the fractional contribution of the inner S\'{e}rsic profile at each radius (Figure~\ref{fig:lightprofiles}). Both S\'{e}rsic profiles were convolved with a Gaussian PSF (FWHM = 0$\farcs$07) before computing the fractions to account for the PSF of STIS. 

\par We find that modeling our data with just two SSPs produces a transition in overall population parameters with radius that is relatively consistent with the transition expected based on the S\'{e}rsic profiles alone. The YMR fractions corresponding to our data show a more gradual transition than the S\'{e}rsic profiles which likely indicates that either the true populations within the galaxy involve more than two SSPs or the S\'{e}rsic profile fits are slightly incorrect. Additionally, because NSCs are often smaller in bluer filters \citep{Carson2015}, the shape discrepancy may be due to the filter (F547M) used to deconstruct the SB profile \citep{Nguyen2018}, which is redder than the mean wavelength of our spectra. Regardless, the agreement seen in this test suggests the observed transition between the two broad stellar populations found in our SED fitting correspond to the two structural components of the NSC from \citet{Nguyen2018}. However, the maps of \citet{Mitzkus2017} in Fig.~\ref{fig:maps} show that at larger radii the dominant stellar population must be more similar to the inner component, but this may be due in part to their lower resolution.  

\subsection{Impact of BHB Stars} \label{sec:BHB}

\par The mean population parameters at radii beyond 0$\farcs$25 indicate the presence of very metal-poor populations with ages $\sim$1~Gyr. Because such young metal-poor populations are rarely observed, we investigated if the presence of BHB stars in the outer parts of the NSC would result in underestimated ages. For instance, \citet{Boecker2020} show that the integrated light coming only from the BHB stars present in the nearby NSC M54 was best-fit by very young ($<$ 1~Gyr) SSPs, which further illustrates the possible underestimation if a significant population of BHB stars is present.  As discussed in Section~\ref{sec:SSPmodels}, it is common to observe populations of BHB stars in GCs, and these stars could be present in an NSC that formed via cluster inspiral.

\par
\begin{figure}[htp]
    \centering
    \includegraphics[width=\linewidth]{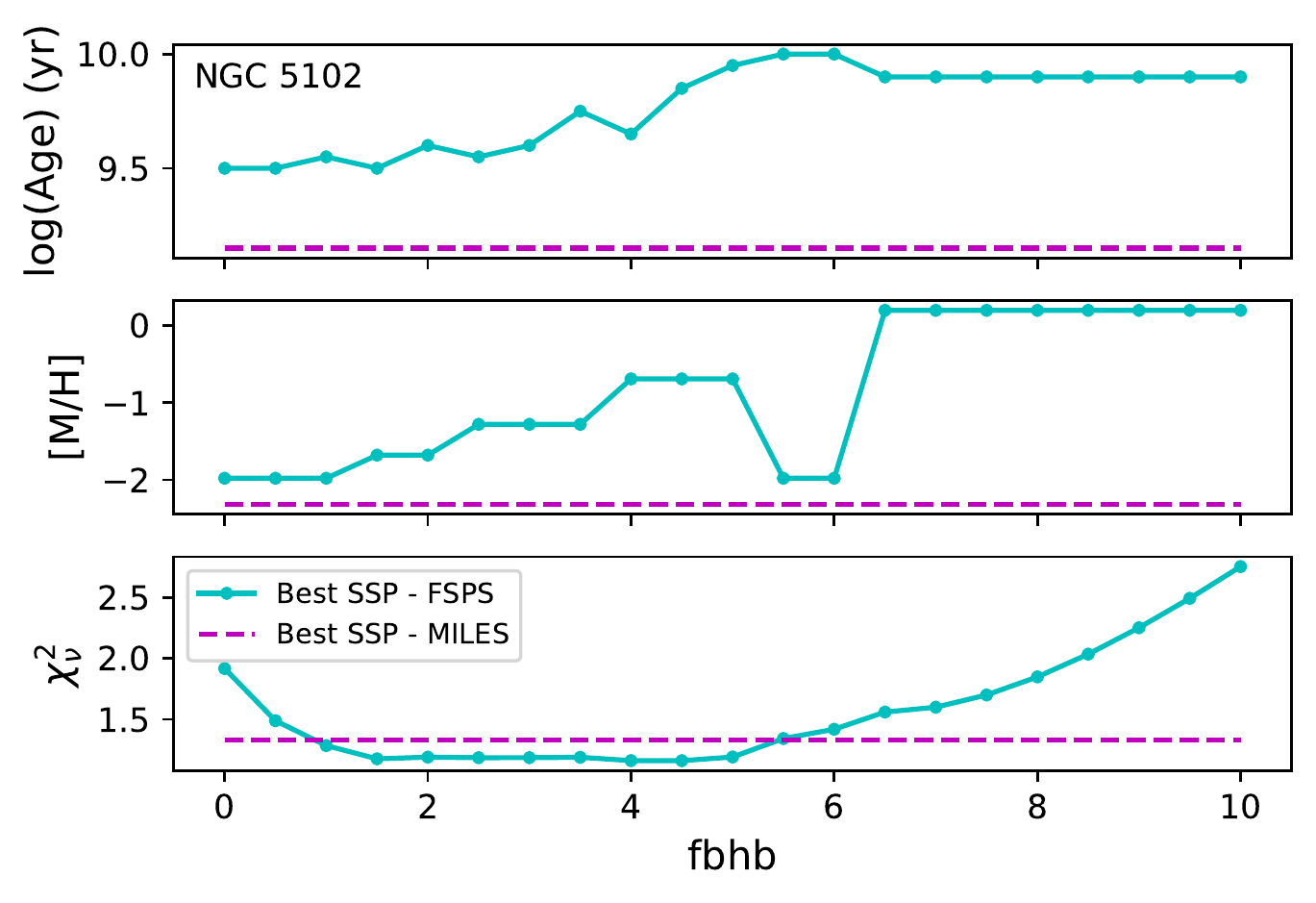}
    \caption{Best-fit log(ages) and metallicities from SSP fits to a spectrum binned over the radial range [0$\farcs$3, 0$\farcs$9] with the MILES model grid (magenta) and FSPS model grids with varying fractions of BHB stars parameterized by \textsc{fbhb} (cyan).  These fits demonstrate that older SSP models hosting large populations of BHB stars provide better fits than the MILES result with similar metallicities.}
    \label{fig:FSPS}
\end{figure}

 To gauge the plausibility that our age measurements could be affected by the presence of BHB stars in the NSC of NGC~5102, SSP fits using both the MILES and FSPS model grids were performed on the binned spectra in the radial range [0$\farcs$3, 0$\farcs$9]. For the SSP fits using the FSPS model grid, only SSPs affected by the \textsc{fbhb} parameter (see Section~\ref{sec:SSPmodels} for details) were used restricting the age range of the FSPS models to [3~Gyr, 14~Gyr]. The results of this test are shown in Figure~\ref{fig:FSPS} where the best-fit SSP parameters from the MILES grid are labeled ``Best SSP - MILES" and the best parameters resulting from the FSPS model grids as a function of \textsc{fbhb} are labeled ``Best SSP- FSPS". At \textsc{fbhb} values between 1 and 5, the bottom panel of Fig.~\ref{fig:FSPS} shows that older SSP models with low metallicities result in better fit qualities than our original MILES best-fit SSP. We note that these high \textsc{fbhb} values are somewhat unphysical in that they suggest a larger population of BHB stars than expected for the given SSP. However, this could be due to the presence of abundance variations in the NSC populations leading to the presence of large numbers of extreme BHB stars with higher temperatures than are used in the FSPS models (which has BHB stars with temperatures $\leq$10,000~K).

\par This result suggests that the age measurements for radii $\geq$ 0$\farcs$3 obtained by fits with the MILES model grid could potentially be impacted by the presence of BHB stars in the NSC. However, a more detailed investigation would be required to verify the extent of the effect.

\section{Conclusions}

\par We present long-slit spectroscopic observations that resolve the NSCs in three galaxies, NGC~5102, NGC~5206, and NGC~205. Using stellar population synthesis we examine the radial gradients of stellar population parameters in these galaxies.  Due to the high signal-to-noise in NGC~5102, it provided the most interesting and significant results. The primary results for this galaxy follow:

\begin{enumerate}
  \item Our composite stellar population fits with MILES SSP models reveal the presence of significant age and metallicity gradients within 0$\farcs$9 of the NSC's center. The very center of the NSC was best-fit by a young (LW age $\sim$400~Myr), metal-rich (LW [M/H] $\sim$-0.4) CSP model. The outermost radii were best-fit by CSP models with older ages (mean LW ages $\sim$1~Gyr) and lower metallicities (mean LW [M/H]'s $\sim$-1.6).  
  \item The presence of a compact, young, and metal-rich population at the very center of the NSC suggests a period of \textit{in situ} formation in this region. 
  \item At larger radii within NGC~5102's NSC, the stellar populations are older and more metal-poor than those in the center.  These populations are also significantly more metal-poor than those of the surrounding galaxy \citep{Mitzkus2017}.  These observations suggest formation from cluster inspiral. 
  \item We find evidence linking the observed transition between the two broad stellar populations in this work to the S\'{e}rsic fits of the SB profile of NGC~5102 found by \cite{Nguyen2018}. The two S\'{e}rsic profiles corresponding to the NSC transition at $\sim$0$\farcs$2 which corresponds closely to the transition radius in population parameters derived from our fits. Furthermore, when modelling our data with two MILES SSP models to represent each broad population, the fractional contributions of each model closely resemble the fractional contributions from each S\'{e}rsic profile at the corresponding radii. 
  \item SSP fits to the binned spectra in the radial range [0$\farcs$3, 0$\farcs$9] with FSPS models of varying BHB fractions suggest that the age of the young metal-poor population measured with the MILES model grid could be underestimated if enough BHB stars are present.  
\end{enumerate}

\par We also derived integrated age and metallicity measurements for the NSCs in NGC~5206 and NGC~205 using integrated light spectra of each galaxy (integrated out to the maximum useful radius). The NSC of NGC~5206 was found to be the oldest cluster in our sample with a mean LW age $\sim$5.8~Gyr and [M/H] $\sim$-0.72 from regularized CSP fits. NGC~205 was found to be rather young with a mean LW age $\sim$0.73~Gyr and [M/H] $\sim$-0.78 (also based on regularized CSP fits). Additionally, we detect a mild metallicity gradient in the NSC of NGC~5206 that is similar to the gradient present in NGC~5102 but with lower significance.

\section*{Acknowledgements}
We thank Martin Mitzkus for sharing his data, and Charlie Conroy for help with FSPS. CHH and ACS acknowledge support from a grant associated with HST program GO-14742.  
    
\bibliography{references}
\bibliographystyle{aasjournal}

\appendix
\restartappendixnumbering
\section{NGC 5206} \label{app:5206}

\par
\begin{figure*}[!b]
    \begin{minipage}{0.45\textwidth}
        \includegraphics[width=\linewidth]{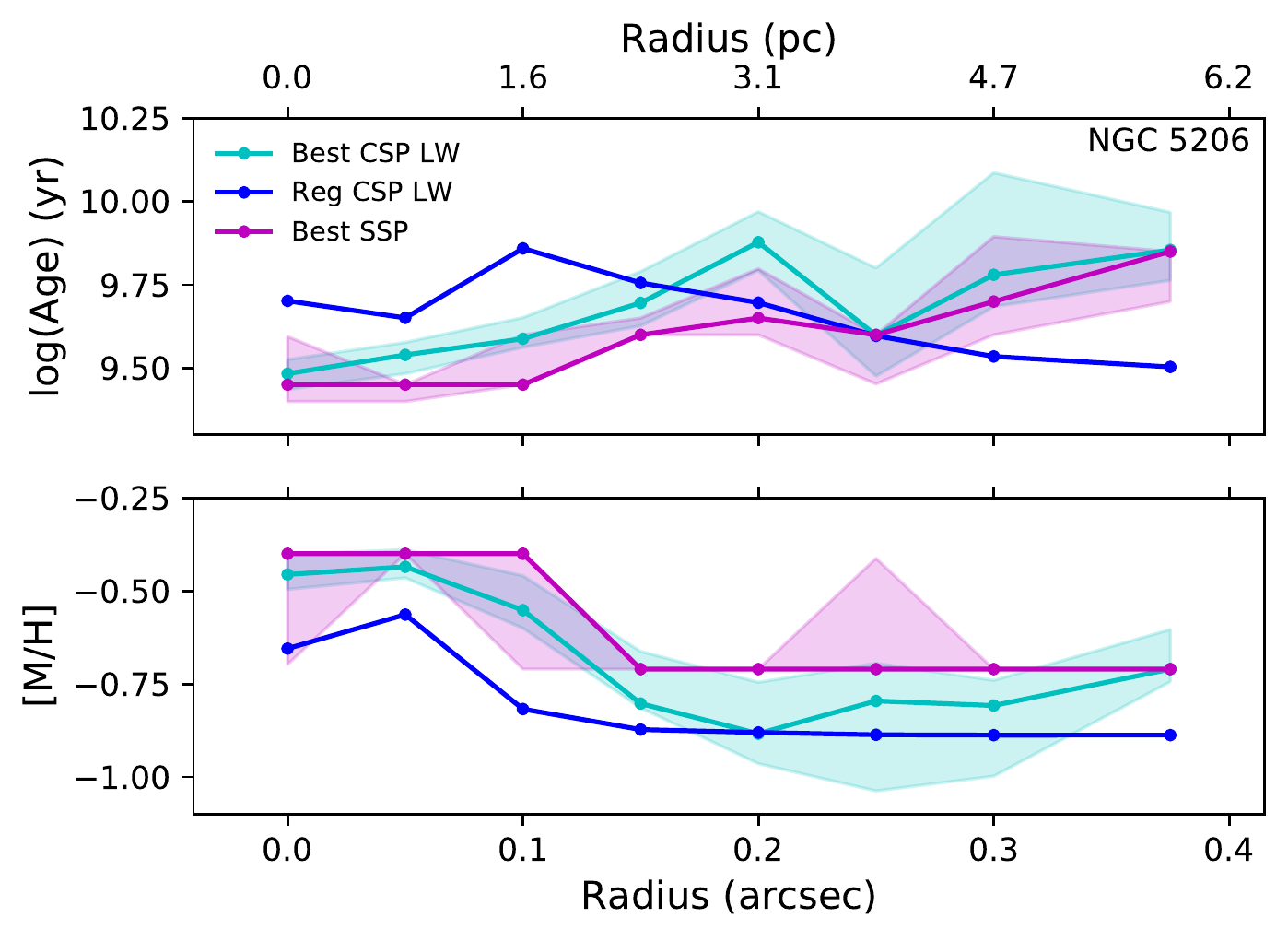}
    \end{minipage} \quad
    \begin{minipage}{0.45\textwidth}                             \includegraphics[width=\linewidth]{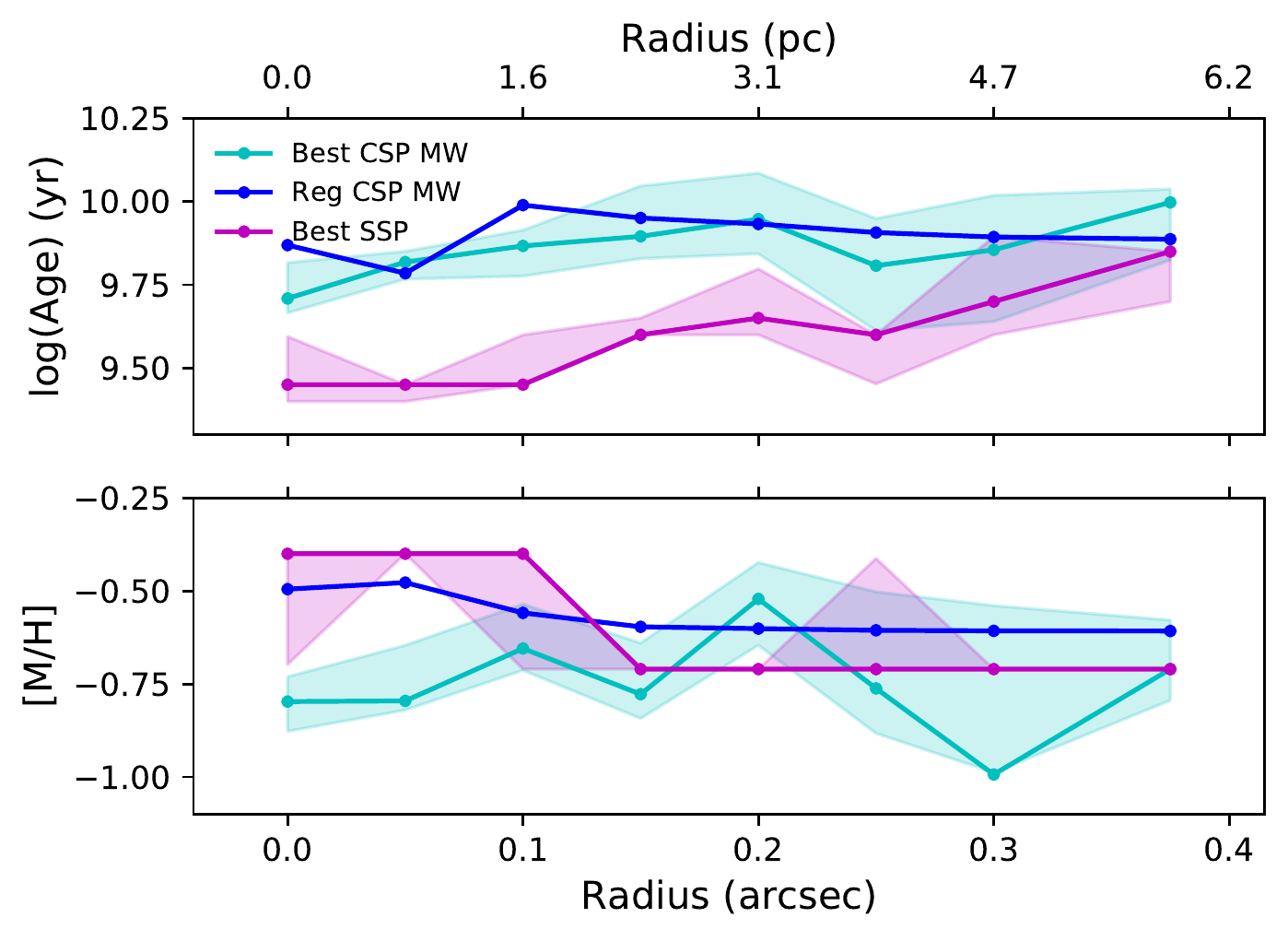}
    \end{minipage}
\caption{The best-fit log(ages) and metallicities as functions of radius for NGC~5206. Left: light-weighted unregularized CSP fits are shown in cyan with the shaded region giving the MC simulation 16-th and 84-th percentiles.  Light-weighted regularized CSP fit solutions are shown in blue, and best SSP fit solutions are shown in magenta with the shaded 16-th and 84-th percentiles from MC simulations. Right: Same as the left panel, but with mass-weighted CSP parameters.}
\label{fig:bestparams5206}
\end{figure*}

 Our resolved stellar population synthesis results for NGC~5206 are graphically presented in Figure~\ref{fig:bestparams5206} and listed numerically in Table~\ref{tab:NGC5206_Results_ALL}. These results reveal no distinct age gradients within the NSC. However, our results suggest that a mild metallicity gradient may be present in NGC~5206.  We note this gradient is not very significant, and higher quality data would be needed to verify this trend. The narrow radial range explored for this galaxy is a direct result of poor data quality -- the data's $S/N$ does not surpass our binning threshold of 10 at larger radii. Furthermore, the maximum regularization values determined for this data at each radius are comparable to the largest regularization values given for NGC~5102. These high regularization parameters result in excessively smoothed distributions of weights indicating that the stellar population synthesis results obtained from this data are not very constraining.

\input{NGC5206_TABLE}
\newpage

\restartappendixnumbering

\section{NGC 205} \label{app:205}

\par The resolved stellar population synthesis results for NGC~205 can be found in Figure~\ref{fig:bestparams5206} as well as Table~\ref{tab:NGC205_Results_ALL}. These results do not reveal notable age or metallicity gradients within the NSC. The low $S/N$ values of the data for this galaxy also dictate the small radial range explored here. As in NGC~5206, we find very high regularization values that provide minimal constraints on the stellar populations of the NGC~205 NSC. 

\begin{figure*}[b]
    \begin{minipage}{0.48\textwidth}
        \includegraphics[width=\linewidth]{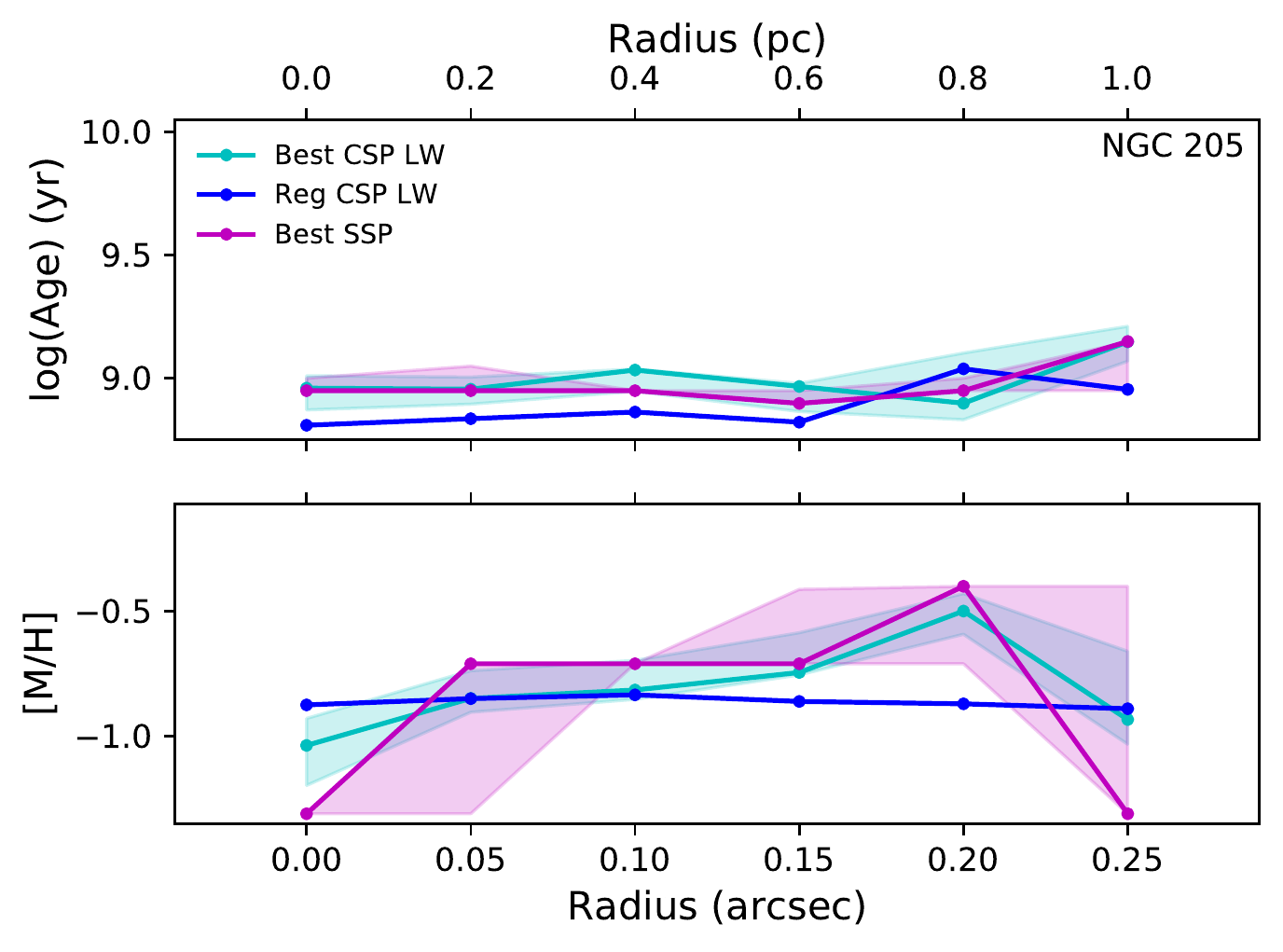}
    \end{minipage} \quad
    \begin{minipage}{0.48\textwidth}                                                   \includegraphics[width=\linewidth]{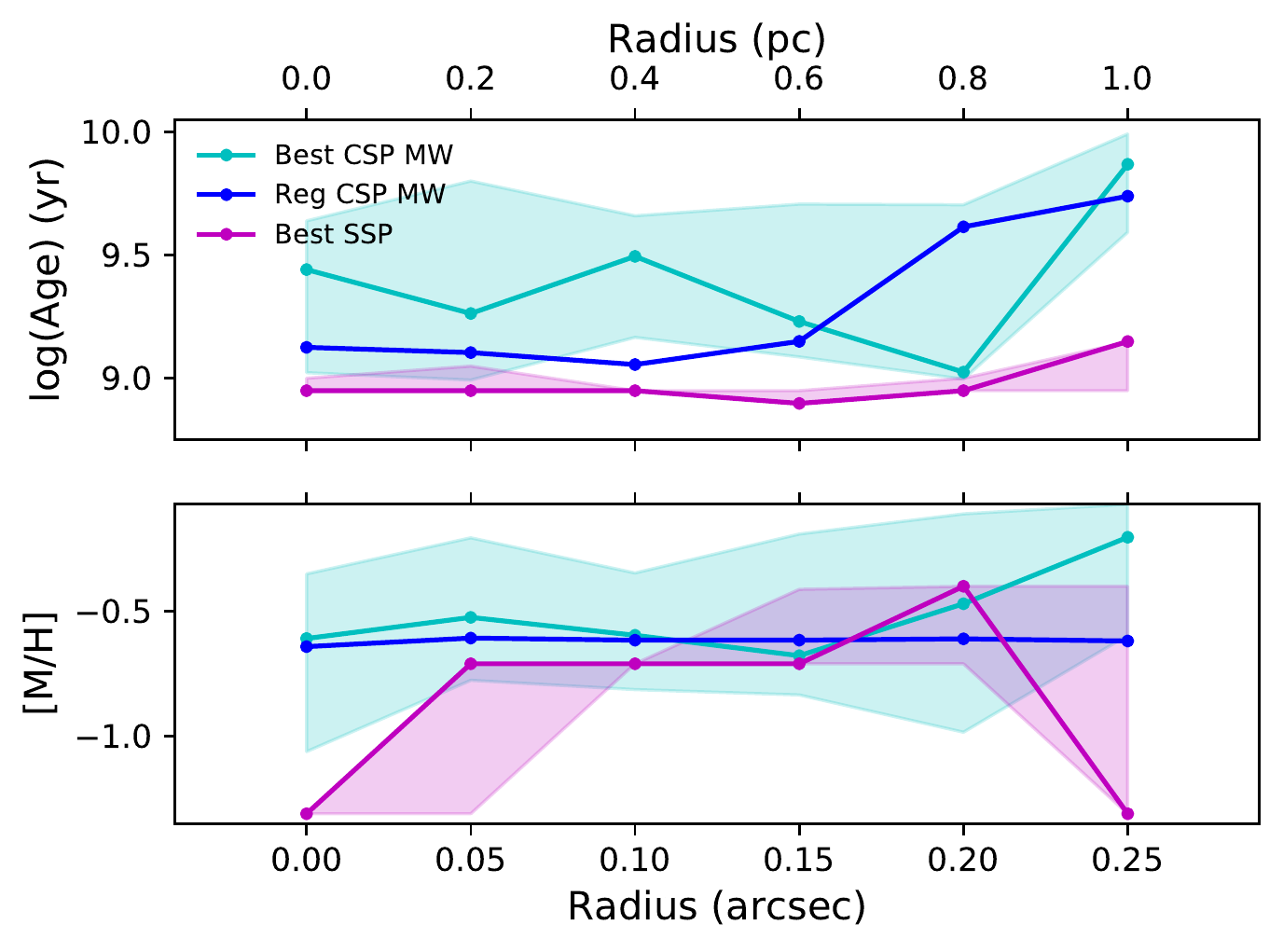}
    \end{minipage}
\caption{Identical to Figure~\ref{fig:bestparams5206} but for NGC~205.}
\label{fig:bestparams205}
\end{figure*}

\input{NGC205_TABLE}


\end{document}

%% file: GAL_PROPERTY_TABLE.tex
\begin{deluxetable*}{cccccccccc}
\tabletypesize{\scriptsize}
\tablecolumns{10}
\tablecaption{\label{tab1} Galaxy Sample and NSC Properties}
\tablehead{
\colhead{Object} & \colhead{$\alpha$ (J2000)} & \colhead{$\delta$ (J2000)} & \colhead{Distance} & \colhead{Galaxy Stellar Mass$^{(3)}$} & \colhead{NSC Mass$^{(3)}$} & \colhead{NSC r$_\textrm{eff}$ $^{(4)}$} & \colhead{NSC $\sigma$ $^{(5)}$} & \colhead{V$_\textrm{r}$} & \colhead{Galaxy M$_\textrm{B}$ $^{(9)}$}\\[-0.2cm]
\colhead{} & \colhead{[h m s]} & \colhead{[$^{\circ}$ $'$ $''$]} & \colhead{[Mpc]} & \colhead{[$\times 10^9$ M$_\odot$]} & \colhead{[$\times 10^7$ M$_\odot$]} & \colhead{[arcsec]} & \colhead{[km/s]} & \colhead{[km/s]} & \colhead{[mag]}}
\startdata
\hline
NGC 5102 & 13:21:57.6 & -36:37:48.9 & 3.7 $^{(1)}$ & 6.9 & 7.30 $\pm$ 2.34 & 1.6 $\pm$ 0.1 & 50 & 468 $\pm$ 2 $^{(6)}$ & -17.9 $\pm$ 0.2\\ 
NGC 5206 & 13:33:44.0 & -48:09:04.2 & 3.2 $^{(1)}$ & 2.5 & 1.54 $\pm$ 0.51 & 0.5 $\pm$ 0.1 & 40 & 571 $\pm$ 10 $^{(7)}$ & -16.1 $\pm$ 0.6\\
NGC 205 & 00:40:22.1 & +41:41:07.1 & 0.8 $^{(2)}$ & 1.1 & 0.20 $\pm$ 0.10 & 0.3 $\pm$ 0.1 & 20 & -241 $\pm$ 3 $^{(8)}$ & -16.0 $\pm$ 0.2\\
\hline
\enddata
\tablecomments{$^{(1)}$ \cite{Tully2015}, $^{(2)}$ \cite{Tully2013}, $^{(3)}$ Masses were derived from dynamically measured M/L ratios \cite{Nguyen2018}, $^{(4)}$ \cite{Nguyen2018}, $^{(5)}$ NSC velocity dispersions used in this work estimated from measurements near the radii explored here \cite{Nguyen2018}, $^{(6)}$ \cite{Koribalski2004}, $^{(7)}$ \cite{Cote1997}, $^{(8)}$ \cite{Bender1991}, $^{(9)}$ HyperLeda  http://leda.univ-lyon1.fr/}
\end{deluxetable*}

%% file: NGC5102_TABLE.tex
\begin{deluxetable*}{cccccccccccc}
\tablecolumns{12}
\tabletypesize{\scriptsize}
\tablecaption{NGC 5102 Fit Results \label{tab:NGC5102_Results_ALL}}
\tablehead{
    \colhead{} & \colhead{LW CSP} & \colhead{LW Reg CSP} & \colhead{MW CSP} & \colhead{MW Reg CSP} & \colhead{SSP} & \colhead{LW CSP} & \colhead{LW Reg CSP} & \colhead{MW CSP} & \colhead{MW Reg CSP} & \colhead{SSP} & \colhead{} \\[-0.2cm]
    \colhead{Radius} & \colhead{log(Age)} & \colhead{log(Age)} & \colhead{log(Age)} & \colhead{log(Age)} & \colhead{log(Age)} & \colhead{[M/H]} & \colhead{[M/H]} & \colhead{[M/H]} & \colhead{[M/H]} & \colhead{[M/H]} & \colhead{\textsc{regul}}
}

\startdata
\hline
0$\farcs$00 & 8.68$^{+0.02}_{-0.00}$ & 8.65 & 8.69$^{+0.11}_{-0.01}$ & 8.66 & 8.70$^{+0.05}_{-0.05}$ & -0.37$^{+0.08}_{-0.05}$ & -0.24 & -0.36$^{+0.10}_{-0.03}$ & -0.25 & -0.40$^{+0.00}_{-0.00}$ & 12\\ 
0$\farcs$05 & 8.68$^{+0.01}_{-0.01}$ & 8.66 & 8.68$^{+0.07}_{-0.00}$ & 8.67 & 8.70$^{+0.00}_{-0.05}$ & -0.47$^{+0.06}_{-0.01}$ & -0.35 & -0.42$^{+0.02}_{-0.13}$ & -0.33 & -0.40$^{+0.00}_{-0.00}$ & 20\\ 
0$\farcs$10 & 8.71$^{+0.01}_{-0.01}$ & 8.70 & 8.71$^{+0.01}_{-0.01}$ & 8.71 & 8.75$^{+0.00}_{-0.05}$ & -0.56$^{+0.09}_{-0.01}$ & -0.43 & -0.49$^{+0.06}_{-0.02}$ & -0.42 & -0.40$^{+0.00}_{-0.00}$ & 14\\ 
0$\farcs$15 & 8.76$^{+0.02}_{-0.02}$ & 8.77 & 8.78$^{+0.02}_{-0.01}$ & 8.78 & 8.85$^{+0.00}_{-0.00}$ & -0.79$^{+0.14}_{-0.04}$ & -0.77 & -0.70$^{+0.07}_{-0.08}$ & -0.75 & -0.71$^{+0.00}_{-0.00}$ & 16\\ 
0$\farcs$20 & 8.94$^{+0.02}_{-0.03}$ & 8.90 & 9.04$^{+0.05}_{-0.07}$ & 8.97 & 8.90$^{+0.00}_{-0.00}$ & -1.41$^{+0.14}_{-0.10}$ & -1.36 & -1.50$^{+0.15}_{-0.05}$ & -1.40 & -1.71$^{+0.00}_{-0.00}$ & 37\\ 
0$\farcs$25 & 9.00$^{+0.05}_{-0.05}$ & 8.94 & 9.13$^{+0.03}_{-0.02}$ & 9.05 & 8.95$^{+0.00}_{-0.00}$ & -1.68$^{+0.09}_{-0.09}$ & -1.61 & -1.76$^{+0.02}_{-0.06}$ & -1.63 & -1.71$^{+0.00}_{-0.00}$ & 69\\ 
0$\farcs$30 & 8.84$^{+0.03}_{-0.07}$ & 8.82 & 9.00$^{+0.04}_{-0.06}$ & 8.97 & 8.95$^{+0.00}_{-0.00}$ & -1.41$^{+0.06}_{-0.05}$ & -1.47 & -1.40$^{+0.15}_{-0.09}$ & -1.41 & -1.71$^{+0.00}_{-0.00}$ & 76\\ 
0$\farcs$35 & 8.91$^{+0.04}_{-0.07}$ & 8.85 & 9.11$^{+0.03}_{-0.02}$ & 9.08 & 8.95$^{+0.10}_{-0.00}$ & -1.42$^{+0.05}_{-0.04}$ & -1.40 & -1.55$^{+0.05}_{-0.05}$ & -1.41 & -1.71$^{+0.00}_{-0.00}$ & 82\\ 
0$\farcs$40 & 9.06$^{+0.03}_{-0.07}$ & 9.00 & 9.22$^{+0.02}_{-0.03}$ & 9.21 & 9.05$^{+0.10}_{-0.00}$ & -1.59$^{+0.12}_{-0.02}$ & -1.51 & -1.79$^{+0.13}_{-0.08}$ & -1.56 & -1.71$^{+0.00}_{-0.61}$ & 91\\ 
0$\farcs$45 & 9.01$^{+0.03}_{-0.06}$ & 8.94 & 9.18$^{+0.01}_{-0.06}$ & 9.20 & 9.05$^{+0.10}_{-0.05}$ & -1.59$^{+0.17}_{-0.04}$ & -1.56 & -1.73$^{+0.12}_{-0.07}$ & -1.57 & -1.71$^{+0.00}_{-0.61}$ & 128\\ 
0$\farcs$50 & 9.07$^{+0.06}_{-0.09}$ & 8.98 & 9.18$^{+0.04}_{-0.03}$ & 9.33 & 9.15$^{+0.00}_{-0.15}$ & -1.97$^{+0.13}_{-0.07}$ & -1.67 & -2.06$^{+0.10}_{-0.06}$ & -1.72 & -2.32$^{+0.61}_{-0.00}$ & 156\\ 
0$\farcs$55 & 9.08$^{+0.09}_{-0.09}$ & 8.96 & 9.31$^{+0.02}_{-0.06}$ & 9.43 & 9.35$^{+0.00}_{-0.20}$ & -1.66$^{+0.10}_{-0.12}$ & -1.47 & -1.90$^{+0.15}_{-0.05}$ & -1.56 & -2.32$^{+0.00}_{-0.00}$ & 134\\ 
0$\farcs$60 & 9.12$^{+0.07}_{-0.09}$ & 8.93 & 9.33$^{+0.01}_{-0.10}$ & 9.46 & 9.15$^{+0.00}_{-0.00}$ & -1.60$^{+0.10}_{-0.15}$ & -1.34 & -1.90$^{+0.16}_{-0.10}$ & -1.46 & -2.32$^{+0.00}_{-0.00}$ & 208\\ 
0$\farcs$65 & 9.23$^{+0.06}_{-0.08}$ & 8.90 & 9.63$^{+0.14}_{-0.29}$ & 9.56 & 9.35$^{+0.00}_{-0.15}$ & -1.82$^{+0.21}_{-0.10}$ & -1.04 & -1.33$^{+0.47}_{-0.62}$ & -0.93 & -2.32$^{+0.00}_{-0.00}$ & 516\\ 
0$\farcs$70 & 9.08$^{+0.06}_{-0.09}$ & 8.84 & 9.31$^{+0.16}_{-0.06}$ & 9.47 & 9.20$^{+0.15}_{-0.05}$ & -1.58$^{+0.20}_{-0.04}$ & -1.11 & -1.84$^{+0.46}_{-0.18}$ & -1.07 & -2.32$^{+0.00}_{-0.00}$ & 324\\ 
0$\farcs$75 & 9.12$^{+0.05}_{-0.11}$ & 8.91 & 9.28$^{+0.04}_{-0.08}$ & 9.44 & 9.30$^{+0.05}_{-0.15}$ & -1.53$^{+0.17}_{-0.13}$ & -1.12 & -1.74$^{+0.23}_{-0.12}$ & -1.09 & -2.32$^{+0.61}_{-0.00}$ & 379\\ 
0$\farcs$80 & 9.08$^{+0.02}_{-0.05}$ & 9.05 & 9.21$^{+0.03}_{-0.04}$ & 9.53 & 9.25$^{+0.00}_{-0.05}$ & -1.46$^{+0.04}_{-0.08}$ & -1.32 & -1.47$^{+0.04}_{-0.07}$ & -1.30 & -1.71$^{+0.00}_{-0.00}$ & 231\\ 
0$\farcs$85 & 9.30$^{+0.01}_{-0.09}$ & 9.04 & 9.34$^{+0.08}_{-0.06}$ & 9.55 & 9.35$^{+0.00}_{-0.10}$ & -1.93$^{+0.20}_{-0.05}$ & -1.21 & -2.03$^{+0.33}_{-0.10}$ & -1.15 & -2.32$^{+0.61}_{-0.00}$ & 335\\
0$\farcs$90 & 9.24$^{+0.04}_{-0.12}$ & 9.01 & 9.31$^{+0.05}_{-0.03}$ & 9.52 & 9.20$^{+0.10}_{-0.00}$ & -1.74$^{+0.11}_{-0.11}$ & -1.13 & -1.81$^{+0.07}_{-0.11}$ & -1.08 & -1.71$^{+0.40}_{-0.61}$ & 402\\ 
\hline
Integrated & 8.75$^{+0.00}_{-0.01}$ & 8.76 & 8.75$^{+0.04}_{-0.01}$ & 8.76 & 8.85$^{+0.00}_{-0.15}$ & -0.87$^{+0.10}_{-0.04}$ & -0.80 & -0.75$^{+0.03}_{-0.09}$ & -0.74 & -0.71$^{+0.31}_{-0.00}$ & 15\\ 
\hline 
\enddata
\tablecomments{Columns labeled with ``LW" indicate light-weighted quantities, while ``MW" refers to mass-weighted.  Unregularized CSP solutions are indicated with ``CSP" column labels; the ``Reg CSP" column labels refer to regularized CSP fit parameters.
Errors listed in this table were derived by taking the difference between the best fit parameters and the 16-th/84-th percentiles for each quantity resulting from MC simulations (see Section~\ref{sec:cspfits} for more details). The column labeled ``\textsc{regul}" gives the value of the \textsc{ppxf} \textsc{regul} parameter used for regularized fits.}

\end{deluxetable*}

%% file: NGC5206_TABLE.tex
\begin{deluxetable*}{cccccccccccc}
\tablecolumns{12}
\tabletypesize{\scriptsize}
\tablecaption{NGC 5206 Fit Results \label{tab:NGC5206_Results_ALL}}
\tablehead{
    \colhead{} & \colhead{LW CSP} & \colhead{LW Reg CSP} & \colhead{MW CSP} & \colhead{MW Reg CSP} & \colhead{SSP} & \colhead{LW CSP} & \colhead{LW Reg CSP} & \colhead{MW CSP} & \colhead{MW Reg CSP} & \colhead{SSP} & \colhead{} \\[-0.2cm]
    \colhead{Radius} & \colhead{log(Age)} & \colhead{log(Age)} & \colhead{log(Age)} & \colhead{log(Age)} & \colhead{log(Age)} & \colhead{[M/H]} & \colhead{[M/H]} & \colhead{[M/H]} & \colhead{[M/H]} & \colhead{[M/H]} & \colhead{\textsc{regul}}
}

\startdata
\hline
0$\farcs$00 & 9.48$^{+0.04}_{-0.05}$ & 9.70 & 9.71$^{+0.11}_{-0.04}$ & 9.87 & 9.45$^{+0.14}_{-0.05}$ & -0.46$^{+0.05}_{-0.04}$ & -0.65 & -0.80$^{+0.07}_{-0.08}$ & -0.50 & -0.40$^{+0.00}_{-0.30}$ & 212\\ 
0$\farcs$05 & 9.54$^{+0.04}_{-0.06}$ & 9.65 & 9.82$^{+0.03}_{-0.05}$ & 9.78 & 9.45$^{+0.00}_{-0.05}$ & -0.44$^{+0.05}_{-0.03}$ & -0.56 & -0.80$^{+0.15}_{-0.02}$ & -0.48 & -0.40$^{+0.00}_{-0.00}$ & 140\\ 
0$\farcs$10 & 9.59$^{+0.06}_{-0.03}$ & 9.86 & 9.87$^{+0.05}_{-0.09}$ & 9.99 & 9.45$^{+0.15}_{-0.00}$ & -0.55$^{+0.09}_{-0.05}$ & -0.82 & -0.65$^{+0.12}_{-0.06}$ & -0.56 & -0.40$^{+0.00}_{-0.31}$ & 946\\ 
0$\farcs$15 & 9.70$^{+0.09}_{-0.07}$ & 9.76 & 9.90$^{+0.15}_{-0.07}$ & 9.95 & 9.60$^{+0.05}_{-0.00}$ & -0.80$^{+0.14}_{-0.01}$ & -0.87 & -0.78$^{+0.14}_{-0.07}$ & -0.60 & -0.71$^{+0.00}_{-0.00}$ & 1831\\ 
0$\farcs$20 & 9.88$^{+0.09}_{-0.08}$ & 9.70 & 9.95$^{+0.14}_{-0.10}$ & 9.93 & 9.65$^{+0.15}_{-0.05}$ & -0.88$^{+0.14}_{-0.08}$ & -0.88 & -0.52$^{+0.10}_{-0.12}$ & -0.60 & -0.71$^{+0.00}_{-0.00}$ & 2232\\ 
0$\farcs$25 & 9.60$^{+0.20}_{-0.12}$ & 9.60 & 9.81$^{+0.14}_{-0.20}$ & 9.91 & 9.60$^{+0.00}_{-0.15}$ & -0.80$^{+0.10}_{-0.24}$ & -0.89 & -0.76$^{+0.26}_{-0.12}$ & -0.61 & -0.71$^{+0.30}_{-0.00}$ & 2207\\ 
0$\farcs$30 & 9.78$^{+0.31}_{-0.10}$ & 9.53 & 9.86$^{+0.16}_{-0.22}$ & 9.89 & 9.70$^{+0.19}_{-0.10}$ & -0.81$^{+0.07}_{-0.19}$ & -0.89 & -0.99$^{+0.45}_{-0.01}$ & -0.61 & -0.71$^{+0.00}_{-0.00}$ & 2462\\ 
0$\farcs$38 & 9.85$^{+0.11}_{-0.09}$ & 9.50 & 10.00$^{+0.04}_{-0.17}$ & 9.89 & 9.85$^{+0.00}_{-0.15}$ & -0.71$^{+0.11}_{-0.03}$ & -0.89 & -0.71$^{+0.13}_{-0.08}$ & -0.61 & -0.71$^{+0.00}_{-0.00}$ & 2384\\ 
\hline
Integrated & 9.63$^{+0.03}_{-0.03}$ & 9.76 & 9.87$^{+0.02}_{-0.06}$ & 9.90 & 9.70$^{+0.00}_{-0.05}$ & -0.61$^{+0.07}_{-0.04}$ & -0.72 & -0.76$^{+0.06}_{-0.03}$ & -0.55 & -0.71$^{+0.00}_{-0.00}$ & 163\\ 
\hline 
\enddata
\tablecomments{Same as Table~\ref{tab:NGC5102_Results_ALL} but for NGC~5206.
}

\end{deluxetable*}

%% file: NGC205_TABLE.tex
\begin{deluxetable*}{cccccccccccc}
\tablecolumns{12}
\tabletypesize{\scriptsize}
\tablecaption{NGC 205 Fit Results \label{tab:NGC205_Results_ALL}}
\tablehead{
    \colhead{} & \colhead{LW CSP} & \colhead{LW Reg CSP} & \colhead{MW CSP} & \colhead{MW Reg CSP} & \colhead{SSP} & \colhead{LW CSP} & \colhead{LW Reg CSP} & \colhead{MW CSP} & \colhead{MW Reg CSP} & \colhead{SSP} & \colhead{} \\[-0.2cm]
    \colhead{Radius} & \colhead{log(Age)} & \colhead{log(Age)} & \colhead{log(Age)} & \colhead{log(Age)} & \colhead{log(Age)} & \colhead{[M/H]} & \colhead{[M/H]} & \colhead{[M/H]} & \colhead{[M/H]} & \colhead{[M/H]} & \colhead{\textsc{regul}}
}

\startdata
\hline
0$\farcs$00 & 8.96$^{+0.05}_{-0.09}$ & 8.81 & 9.44$^{+0.20}_{-0.42}$ & 9.13 & 8.95$^{+0.05}_{-0.00}$ & -1.04$^{+0.11}_{-0.16}$ & -0.87 & -0.61$^{+0.26}_{-0.45}$ & -0.64 & -1.31$^{+0.00}_{-0.00}$ & 729\\ 
0$\farcs$05 & 8.96$^{+0.05}_{-0.06}$ & 8.84 & 9.26$^{+0.54}_{-0.27}$ & 9.10 & 8.95$^{+0.10}_{-0.00}$ & -0.85$^{+0.11}_{-0.06}$ & -0.85 & -0.52$^{+0.32}_{-0.25}$ & -0.61 & -0.71$^{+0.00}_{-0.60}$ & 718\\ 
0$\farcs$10 & 9.03$^{+0.00}_{-0.09}$ & 8.86 & 9.50$^{+0.16}_{-0.33}$ & 9.06 & 8.95$^{+0.00}_{-0.00}$ & -0.81$^{+0.12}_{-0.04}$ & -0.83 & -0.60$^{+0.25}_{-0.22}$ & -0.62 & -0.71$^{+0.00}_{-0.00}$ & 434\\ 
0$\farcs$15 & 8.97$^{+0.01}_{-0.10}$ & 8.82 & 9.23$^{+0.48}_{-0.14}$ & 9.15 & 8.90$^{+0.05}_{-0.00}$ & -0.74$^{+0.16}_{-0.01}$ & -0.86 & -0.68$^{+0.49}_{-0.16}$ & -0.62 & -0.71$^{+0.30}_{-0.00}$ & 651\\ 
0$\farcs$20 & 8.90$^{+0.20}_{-0.07}$ & 9.04 & 9.02$^{+0.68}_{-0.03}$ & 9.62 & 8.95$^{+0.05}_{-0.00}$ & -0.50$^{+0.07}_{-0.09}$ & -0.87 & -0.47$^{+0.36}_{-0.51}$ & -0.61 & -0.40$^{+0.00}_{-0.31}$ & 782\\ 
0$\farcs$25 & 9.15$^{+0.06}_{-0.08}$ & 8.95 & 9.87$^{+0.12}_{-0.28}$ & 9.74 & 9.15$^{+0.00}_{-0.20}$ & -0.93$^{+0.27}_{-0.10}$ & -0.89 & -0.20$^{+0.13}_{-0.39}$ & -0.62 & -1.31$^{+0.91}_{-0.00}$ & 3132\\ 
\hline
Integrated & 9.06$^{+0.01}_{-0.04}$ & 8.87 & 9.65$^{+0.05}_{-0.10}$ & 8.97 & 8.90$^{+0.00}_{-0.00}$ & -0.85$^{+0.05}_{-0.06}$ & -0.80 & -0.49$^{+0.08}_{-0.09}$ & -0.62 & -0.40$^{+0.00}_{-0.00}$ & 212\\ 
\hline 
\enddata
\tablecomments{Same as Table~\ref{tab:NGC5102_Results_ALL} but for NGC~205.}

\end{deluxetable*}